\documentclass[10pt,journal,twocolumn,twoside]{IEEEtran}

\usepackage{cite}
\ifCLASSINFOpdf
  \usepackage[pdftex]{graphicx}
  \usepackage{epstopdf}
  \graphicspath{{figures/}}
\else
\fi
\usepackage[cmex10]{amsmath}
\usepackage{array}
\usepackage{url}
\usepackage{mdwmath}
\usepackage{mdwtab}

\usepackage{amssymb}
\usepackage{bm}

\usepackage{color}

\usepackage{enumerate}
\usepackage{mathtools}
\usepackage{stfloats}
\usepackage{microtype}
\usepackage[caption=false,font=footnotesize]{subfig}

\begin{document}
\title{Estimation and Mitigation of\\Channel Non-Reciprocity in Massive MIMO}

\author{Orod~Raeesi,~\IEEEmembership{Student~Member,~IEEE,}
	Ahmet~Gokceoglu,~\IEEEmembership{Member,~IEEE,}\\
	and~Mikko~Valkama,~\IEEEmembership{Senior Member,~IEEE}% <-this % stops a space
	\thanks{O. Raeesi, A. Gokceoglu, and M. Valkama are  with the Department of Electronics and Communications Engineering, Tampere University of Technology, Tampere 33720, Finland (e-mail: orod.raeesi@tut.fi; ahmet.gokceoglu@tut.fi; mikko.e.valkama@tut.fi).}% <-this % stops a space
	\thanks{This work was supported by the Finnish Funding Agency for Technology and Innovation (Tekes) under the project ``$5$th Evolution Take of Wireless Communication Networks (TAKE-$5$)'', by the Academy of Finland under the projects $284694$ and $288670$ and TUT Graduate School.}\thanks{This work has been submitted to IEEE for possible publication. Copyright may be transferred without notice, after which this version may no longer be accessible.}}


\markboth{IEEE Transactions on Signal Processing}%
{Mitigation of Channel Non-Reciprocity in Massive MIMO Systems}



\maketitle

\begin{abstract}
Time-division duplex (TDD) based massive MIMO systems rely on the reciprocity of the wireless propagation channels when calculating the downlink precoders based on uplink pilots. However, the effective uplink and downlink channels incorporating the analog radio front-ends of the base station (BS) and user equipments (UEs) exhibit non-reciprocity due to non-identical behavior of the individual transmit and receive chains. When downlink precoder is not aware of such channel non-reciprocity (NRC), system performance can be significantly degraded due to NRC induced interference terms. In this work, we consider a general TDD-based massive MIMO system where frequency-response mismatches at both the BS and UEs, as well as the mutual coupling mismatch at the BS large-array system all coexist and induce channel NRC. Based on the NRC-impaired signal models, we first propose a novel iterative estimation method for acquiring both the BS and UE side NRC matrices and then also propose a novel NRC-aware downlink precoder design which utilizes the obtained estimates. Furthermore, an efficient pilot signaling scheme between the BS and UEs is introduced in order to facilitate executing the proposed estimation method and the NRC-aware precoding technique in practical systems. Comprehensive numerical results indicate substantially improved spectral efficiency performance when the proposed NRC estimation and NRC-aware precoding methods are adopted, compared to the existing state-of-the-art methods.
\end{abstract}


% \fi
\IEEEpeerreviewmaketitle

\section{Introduction} \label{sec:introduction}
\IEEEPARstart{M}{assive} MIMO is one of the key potential technologies for upcoming $5$G systems \cite{boccardi_five_2014} where base stations (BSs) deploy very large antenna arrays, e.g., several tens or hundreds of antenna units per array, to facilitate high beamforming and spatial multiplexing gains. In such systems, it is not feasible to transmit downlink pilots from each BS antenna in order to estimate the corresponding spatial channels at user equipments (UEs) and feedback the channel state information (CSI) to BS, as the amount of overhead in such approach is proportional to the number of antennas in the BS side \cite{shepard_argos:_2012}. Massive MIMO systems are thus envisioned to primarily deploy time-division duplex (TDD) based radio access and rely on the reciprocity of the physical uplink and downlink channels when obtaining CSI at BS. This, in turn, requires substantially smaller pilot or reference signal overhead being only proportional to the number of UEs \cite{Larsson_TDD_for_massiveMIMO}.

While it is a common assumption in TDD systems that the physical propagation channels are reciprocal within a coherence interval \cite{shepard_argos:_2012,Larsson_TDD_for_massiveMIMO}, the impacts of the BS and UE side transceiver analog front-ends on the \textit{effective downlink and uplink channels} are not reciprocal. This hardware induced phenomenon is often referred to as the channel non-reciprocity (NRC) problem \cite{bourdoux_non-reciprocal_2003, zou_analysis_2014}. Typically, the mismatches in the frequency-responses (FRs) of both the BS and UE side radio front-ends at transmit and receive modes are seen as the main cause of NRC. Another source of NRC considered in literature is the differences in mutual coupling (MC) of BS antenna units and associated RF transceivers under transmit and receive modes \cite{petermann_multi-user_2013, Wei_mutual_coupling}.

The impacts of the NRC on the achievable system performance have been studied in various works in the recent literature. To this end, \cite{zou_analysis_2014} provides downlink sum-rate analysis for a general multi-user MIMO system with zero-forcing (ZF) and eigen-beamforming types of precoding under NRC due to FR mismatch. Then, specifically focusing on massive MIMO systems, \cite{zhang_large-scale_2015, athley_analysis_2015} study achievable downlink sum-rates for maximum-ratio transmission (MRT) and ZF precoding schemes, demonstrating significant performance degradation under practical values of the NRC parameters.

There is also a large amount of work reported in the literature addressing the estimation and mitigation of NRC in TDD based MIMO systems \cite{bourdoux_non-reciprocal_2003, petermann_multi-user_2013, Rogalin_rel_calib, vieira_reciprocity_2014, wei_mutual_2016, shepard_argos:_2012, wei_tdd_2015, guillaud_practical_2005, zou_efficient_2014, guillaud_towards_2013}. These studies can be divided into three main categories as follows: 
\begin{enumerate}[i)]
	\item \label{sec:novelty1} BS carries out ``self-calibration'' using a reference antenna with the help of additional circuitry \cite{bourdoux_non-reciprocal_2003, petermann_multi-user_2013}. This method is capable of estimating BS side NRC only.
	\item BS carries out ``self-calibration'' without additional circuitry. Mutual coupling between antennas is utilized when exchanging pilot signals with the reference antenna \cite{Rogalin_rel_calib, vieira_reciprocity_2014, wei_mutual_2016, shepard_argos:_2012, wei_tdd_2015}. Similar to \ref{sec:novelty1}), also this method estimates only the BS side NRC, and also commonly neglects the mutual coupling mismatch.
	\item BS transmits specific pilot signals to UEs and UEs send back the received signals in certain properly precoded forms to facilitate BS side NRC parameter estimation, which is often referred to as over-the-air (OTA) approach \cite{petermann_multi-user_2013, guillaud_practical_2005, zou_efficient_2014, guillaud_towards_2013}.  
\end{enumerate}

In this work, we focus on OTA-based estimation and mitigation of NRC in a multi-user massive MIMO system context deploying MRT or ZF precoding. The novelty and contributions of this paper can be summarized as follows:
\begin{enumerate}
	\item We consider generalized NRC induced by coexisting FR mismatches of all associated radio transceivers at UE and BS sides as well as the mutual coupling mismatches in the BS side large-array antenna system, unlike many of the earlier works that consider only FR mismatch such as \cite{bourdoux_non-reciprocal_2003, vieira_reciprocity_2014, wei_mutual_2016, shepard_argos:_2012, wei_tdd_2015, guillaud_practical_2005, zou_efficient_2014, Rogalin_rel_calib,guillaud_towards_2013}. In this respect, only \cite{petermann_multi-user_2013} reports similar modeling, however, the proposed mitigation scheme in \cite{petermann_multi-user_2013} is suitable mainly for small scale MIMO systems, e.g., $2$-$4$ BS antennas.
	\item We address the estimation and mitigation of the NRC sources of both the UE and BS sides, unlike many other works that address only BS side NRC, e.g., \cite{shepard_argos:_2012, bourdoux_non-reciprocal_2003, Rogalin_rel_calib, vieira_reciprocity_2014, wei_mutual_2016, wei_tdd_2015, zou_efficient_2014}. As shown in \cite{our_submitted_analysis_journal}, with popular assumption of not having downlink demodulation pilots, UE side NRC can be a major cause of performance degradation in multi-user massive MIMO systems, thus strongly motivating to incorporate such effects in the NRC estimation and mitigation processes.
	\item Unlike other massive MIMO NRC mitigation works \cite{shepard_argos:_2012, Rogalin_rel_calib, vieira_reciprocity_2014, wei_mutual_2016, wei_tdd_2015} which all assume the availability of downlink pilots in the UE side, we consider the appealing massive MIMO scenario in which there are no downlink pilots and thus UEs rely on the statistical properties of the beamformed channels to decode the received downlink signals \cite{yang_performance_2013, zhang_large-scale_2015, Jose_statistical_DL_1, Jose_statistical_DL_2,Larsson_no_DL_pilot}.
	\item We demonstrate the performance of our proposed scheme under imperfect uplink CSI, unlike other works which commonly rely on the perfect uplink CSI assumption \cite{bourdoux_non-reciprocal_2003, vieira_reciprocity_2014, wei_mutual_2016, wei_tdd_2015, zou_efficient_2014}. 
\end{enumerate}

The organization of the paper is as follows. Fundamental signal models of the considered massive MIMO system with MRT and ZF-based precoding schemes under NRC are first presented in Section \ref{sec:system_model}. Then, the NRC-aware downlink precoding approach is formulated for given NRC estimates. In Section \ref{sec:estimation_mitigation}, novel pilot signaling method between the BS and UEs is introduced which is followed by the proposed novel iterative estimation of BS and UE side NRC matrices. The results of empirical performance evaluations in terms of the achievable system spectral efficiency are presented in Section \ref{sec:simulations}, incorporating the proposed estimation-mitigation scheme together with existing state-of-the-art NRC estimation/mitigation methods for reference. Finally, conclusions are drawn in Section \ref{sec:conclusions}.

\textit{Notations:} Throughout the paper, vectors and matrices are denoted with lower and upper case bold letters, respectively, e.g., vector $\mathbf{x}$, matrix $\mathbf{Y}$. The superscripts $\left(.\right)^*$, $\left(.\right)^\mathrm{T}$, $\left(.\right)^\mathrm{H}$, and $\left(.\right)^\dagger$ indicate complex-conjugation, transposition, Hermitian-transpose, and Moore-Penrose pseudo inverse operations, respectively. Expectation operator is shown by $\mathbb{E}{\left[.\right]}$, while $\mathrm{Tr}\left(.\right)$ represents the trace operator. $\mathrm{diag}\left(.\right)$ operator transforms a vector $\mathbf{v}$ to a diagonal matrix with the elements of  $\mathbf{v}$ at its diagonal, and vice versa, reads the diagonal elements of the input matrix into a column vector. $\Re\{.\}$ and $\Im\{.\}$ work element-wise and return real and imaginary parts of complex-valued arguments, respectively. The element in the $i$'th row and $j$'th column of matrix $\mathbf{V}$ is represented by $v_{ij}$, whereas the $i$'th element on the main diagonal of a diagonal matrix $\mathbf{C}$ is shown by $c_{i}$. The complex-valued zero-mean circularly symmetric Gaussian distribution with variance $\sigma^2$ is denoted as $\mathcal{CN}\left(0,\sigma^2\right)$. Finally, $\mathbf{I}_n$ and $\mathbf{0}_n$ denote the $n\times n$ identity and all-zero matrices, respectively.

\section{System Model and Problem Formulation} \label{sec:system_model}
We consider a TDD based single-cell multi-user downlink transmission scenario where the BS with a large number of antenna units, denoted by $N$, transmits to $K$ single-antenna UEs simultaneously in the same time-frequency resource, where $N\gg K$. All signal and system models are written for an arbitrary subcarrier of the underlying orthogonal frequency division multiplexing/multiple access (OFDM/OFDMA) waveform, that is, before IFFT and after FFT on the TX and RX sides, respectively.

In an ideal TDD massive MIMO system, the effective uplink and downlink channels consist of only the reciprocal physical channels. Building on that, the downlink transmission is done by beamforming the multi-user downlink data based on the estimated channels from uplink pilot sequences of length $\tau_u$ symbols \cite{shepard_argos:_2012,Larsson_TDD_for_massiveMIMO}. In this work, we assume the same procedure for the downlink transmission, however, we consider more generalized uplink and downlink effective channel models which are \textit{non-reciprocal} due to radio front-end mismatches and non-idealities. In this respect, the uplink model for channel estimation phase \cite{yang_performance_2013} and the corresponding downlink received signal model in beamformed data transmission phase under the non-reciprocal effective channels can be expressed as
\begin{equation} \label{eq:system_model}
\begin{aligned} 
&\text{Uplink Training}: \quad &&\mathbf{Y}=\sqrt{\rho_{u}\tau_u}\mathbf{G}+\mathbf{Z}_u\\
&\text{Downlink Transmission}: \quad &&\mathbf{r}=\sqrt{\rho_{d}}\mathbf{H}\mathbf{x}+\mathbf{z}_d,
\end{aligned}
\end{equation}
where $\mathbf{x}$ denotes the precoded user data, whereas $\mathbf{G}$ and $\mathbf{H}$ are the effective non-reciprocal uplink and downlink multi-user MIMO channels, respectively, which will be elaborated in detail later in Section \ref{sec:UL DL channels}. $\mathbf{Z}_u$ is the processed noise matrix at the BS, while $\mathbf{z}_d$ denotes the UE side multi-user thermal noise vector, both assumed to consist of i.i.d. $\mathcal{CN}\left(0,1\right)$ elements. The average signal to noise ratios (SNRs) in the uplink and downlink are denoted as $\rho_u$ and $\rho_d$, respectively. This basic system framework is largely based on and following the seminal work by Marzetta in \cite{yang_performance_2013,HienNgo_massMIMO_DLpilot} where reciprocal channels were assumed.

\subsection{Effective and Relative Uplink and Downlink Channels} \label{sec:UL DL channels}
\begin{figure}[!t]
	\centering
	\subfloat[]{\includegraphics[width=0.48\textwidth]{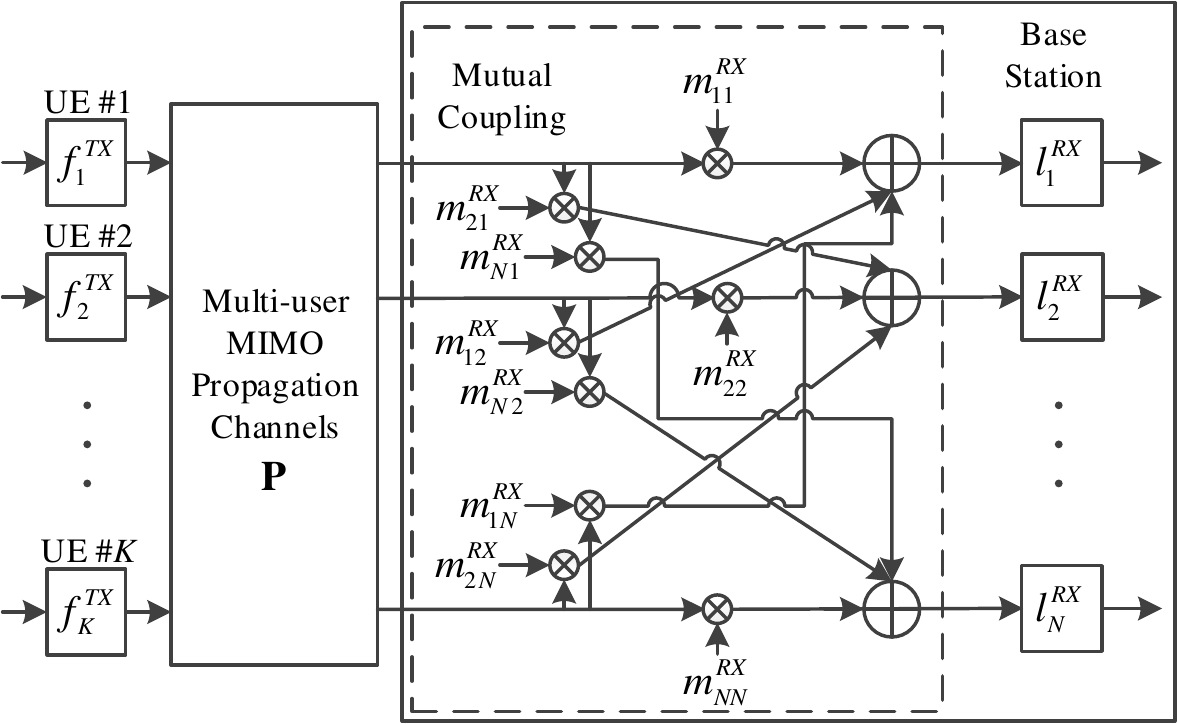}%
		\label{subfig:UL}}
	\hfil
	\subfloat[]{\includegraphics[width=0.48\textwidth]{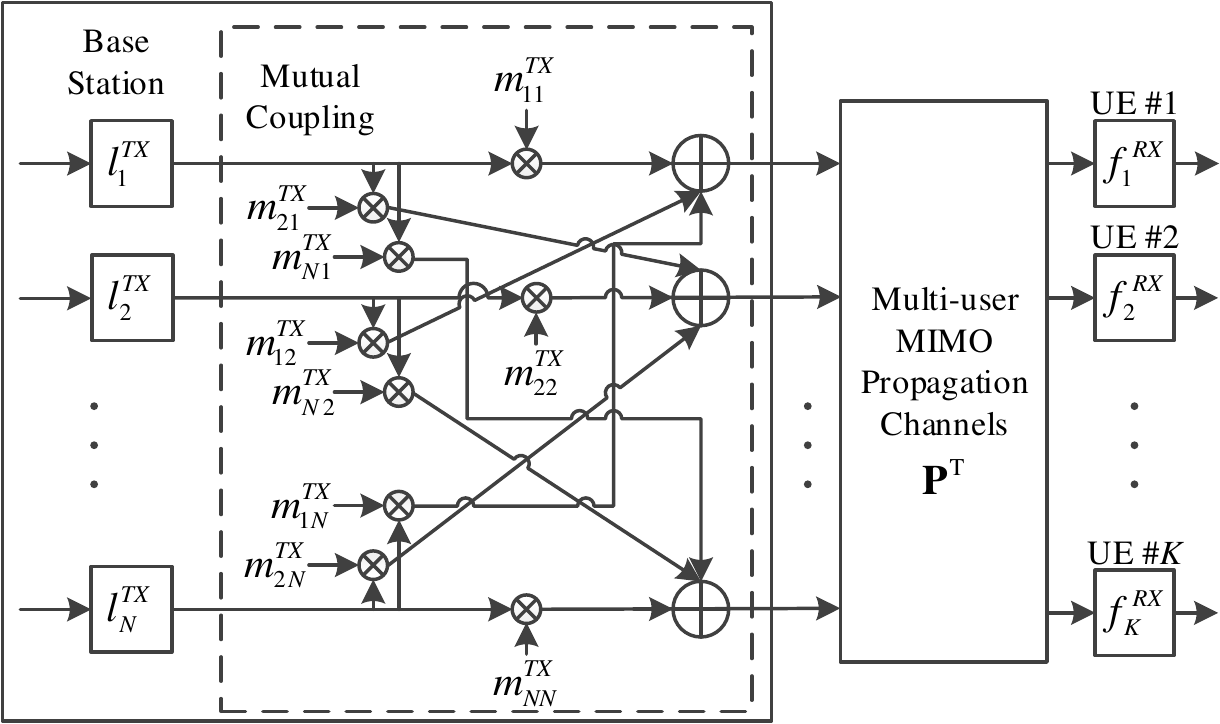}%
		\label{subfig:DL}}
	\caption{Basic system models for \protect\subref{subfig:UL} uplink and \protect\subref{subfig:DL} downlink transmission and reception including physical propagation channels, transceiver frequency responses and antenna mutual coupling in the devices.}
	\label{fig:eff chan}
\end{figure}
As illustrated in \figurename{~\ref{fig:eff chan}}, the complete description of the uplink and downlink effective channels appearing in \eqref{eq:system_model} can be expressed as
\begin{equation} \label{eq:exact_channels}
\begin{aligned} 
\begin{rcases}
\mathbf{G} = \mathbf{E}_{r} \mathbf{P} \mathbf{F}_{t}\\
\mathbf{H} = \mathbf{F}_{r} \mathbf{P}^\mathrm{T} \mathbf{E}_{t} 
\end{rcases} \Rightarrow \mathbf{H} \neq \mathbf{G}^\mathrm{T}, 
\end{aligned} 
\end{equation}
with $\mathbf{E}_{r}= \mathbf{L}_{r} \mathbf{M}_{r}$ and $\mathbf{E}_{t}= \mathbf{M}_{t} \mathbf{L}_{t}$. In above, $\mathbf{F}$ is the joint frequency-response matrix of the UEs, $\mathbf{L}$ is the frequency-response matrix of the BS, $\mathbf{M}$ is the mutual coupling matrix of the BS, $\mathbf{P}$ is the reciprocal physical channel, while the subscripts $t$ and $r$ denote the transmitting and receiving modes, respectively. Note that the frequency-response matrices, $\mathbf{F}$ and $\mathbf{L}$, are diagonal, while the mutual coupling matrix $\mathbf{M}$ in general has both non-zero diagonal and off-diagonal entries.

In general, the effective channels with above assumptions and modeling are clearly non-reciprocal, i.e., $\mathbf{H} \neq  \mathbf{G}^\mathrm{T}$, due to differences in the TX and RX modes of the radio front-end and array responses, i.e., $\mathbf{F}_{t} \neq \mathbf{F}_{r}$, $\mathbf{L}_{t} \neq \mathbf{L}_{r}$ and $\mathbf{M}_{t} \neq \mathbf{M}^\mathrm{T}_{r}$. Hence, the effective uplink and downlink channels can be described relative to each other as 
\begin{equation} \label{eq:relative_channels}
\mathbf{H} = \mathbf{A} \mathbf{G}^\mathrm{T} \mathbf{B},
\end{equation}
%where $\mathbf{G}$ is assumed to have i.i.d. $\mathcal{CN}\left(0,1\right)$ elements. 
where, $\mathbf{A} = \mathbf{F}_{r} \mathbf{F}_{t}^{-1}$ and $\mathbf{B} = \mathbf{L}_{r}^{-1}\left(\mathbf{M}_{r}^{\mathrm{T}}\right)^{-1}\mathbf{M}_{t}\mathbf{L}_{t}$.

In general, $\mathbf{A}$ is a diagonal matrix and the $k$'th diagonal entry, denoted as $a_k$, corresponds to the frequency-response ratio of $k$'th UE at TX and RX modes. In the following, similar to \cite{zou_analysis_2014,petermann_multi-user_2013,zhang_large-scale_2015}, we will use the decomposition of the form $\mathbf{A}=\mathbf{I}_K+\mathbf{A}^{\prime}$, where the diagonal matrix $\mathbf{A}^{\prime}$ measures the deviation from unity frequency-response ratio. The $k$'th diagonal entry of $\mathbf{A}^{\prime}$ is denoted as $a_k^{\prime}$, such that $a_k=1+a_k^{\prime}$.

In \eqref{eq:relative_channels}, $\mathbf{B}$ is a full matrix that incorporates both the frequency-responses and mutual coupling at the BS side. In the following, for notational convenience, we will use the decomposition $\mathbf{B}=\mathbf{I}_N+\mathbf{B}^{\prime}$, where $\mathbf{B}^{\prime}$ accounts for the deviation of diagonal and off-diagonal entries from the ideal reciprocal response.

The detailed modeling of the entries of the above matrices is based on the practical NRC modeling introduced in \cite{petermann_multi-user_2013}, in which $\sigma_\mathrm{F}^2$ is denoting the variance of diagonal elements in $\mathbf{F}_{t}$ and $\mathbf{F}_{r}$, while the corresponding variance of diagonal elements in $\mathbf{L}_{t}$ and $\mathbf{L}_{r}$ is denoted by $\sigma_\mathrm{L}^2$. The power of elements in $\mathbf{M}_{t}$ and $\mathbf{M}_{r}$ is controlled by input reflection coefficients which have the variance $\sigma_\mathrm{M}^2$. 

%Under practical settings, the diagonal entries of $\mathbf{B}$ are close to 1  whereas most of the off-diagonal entries are small. Here, we treat  those off-diagonal entries that are $40$ dB below the strongest diagonal entries as zeros yielding a sparse matrix structure.

The characterization as given in \eqref{eq:exact_channels} and/or \eqref{eq:relative_channels} is generally referred to in literature as channel non-reciprocity \cite{zou_analysis_2014, petermann_multi-user_2013}. The ideal reciprocal channel model is a special case where $\mathbf{A}=\mathbf{B}=\mathbf{I}$, i.e., $\mathbf{A}^{\prime}=\mathbf{B}^{\prime}=\mathbf{0}$.

\subsection{Channel Estimation and Beamforming under NRC}
First, we shortly address the influence of NRC when the downlink transmission is carried out without any processing against the NRC, i.e., NRC-blind precoding is adopted. In this respect, the required downlink channel estimate in BS is obtained from the orthogonal uplink training signals, with the observation model given already on the first line of \eqref{eq:system_model}, complemented, e.g., with LMMSE channel estimator as described in \cite{yang_performance_2013,HienNgo_massMIMO_DLpilot}. This yields formally
\begin{equation} \label{eq:channel_estimate}
\hat{\mathbf{H}}=\hat{\mathbf{G}}^\mathrm{T},
\end{equation}
where $\hat{\mathbf{H}}$ and $\hat{\mathbf{G}}$ are the estimated downlink and uplink effective channels, respectively. %Using minimum mean-square error (MMSE) estimation, the estimated uplink channel can be written in terms of the true uplink channel as
%\begin{equation} \label{eq:channel_estimate}
%\mathbf{H}=\hat{\mathbf{H}}+\bm{\mathcal{E}},
%\end{equation}
%where $\hat{\mathbf{H}}$ and $\bm{\mathcal{E}}$ are independent \cite{MMSE}.}

%as \cite{yang_performance_2013}
%\begin{equation} \label{eq:channel_estimate}
%\hat{\mathbf{H}}=\frac{\sqrt{\rho_u\tau_u}}{1+\rho_u\tau_u}\mathbf{Y}^\mathrm{T},
%\end{equation}

%Then, the estimated channel can be written in terms of the true uplink channel as $\mathbf{H}=\hat{\mathbf{H}}+\bm{\mathcal{E}}$ where $\bm{\mathcal{E}}$ is the zero-mean Gaussian channel estimation noise \cite{HienNgo_massMIMO_DLpilot}.

Using the estimated downlink effective channel in \eqref{eq:channel_estimate}, the user data vector  $\mathbf{s} = \left[s_1,...,s_K\right]^\mathrm{T} \in \mathbb{C}^K$ which is assumed to have element-wise power normalization of the form $\mathbb{E}\left[\left|s_{k}\right|^2\right] = 1$, is precoded as
\begin{equation} \label{eq:precoded_symbols}
\mathbf{x} = \mathbf{U}\mathbf{s},
\end{equation}
where the NRC-blind linear precoding matrix $\mathbf{U}$ reads \cite{HienNgo_massMIMO_DLpilot}
\begin{equation} \label{eq:MRT_ZF}
\begin{aligned}
\mathbf{U}=\begin{cases}\beta \hat{\mathbf{H}}^\mathrm{H},\quad &\text{for MRT}\\
\beta \hat{\mathbf{H}}^\mathrm{H}\left(\hat{\mathbf{H}}\hat{\mathbf{H}}^\mathrm{H}\right)^{-1},\quad &\text{for ZF}.
\end{cases}
\end{aligned}
\end{equation}
In above, without loss of generality, the scalar $\beta$ can be chosen to satisfy unit average transmit power constraint as \cite{yang_performance_2013}
%\begin{equation} \label{eq:beta}
%\begin{aligned}
%\beta=\begin{cases} \sqrt{\frac{\tau_u \rho_u + 1}{N K \tau_u \rho_u}},\quad &\text{for MRT}\\
%\sqrt{\frac{\left(N-K\right)\tau_u \rho_u}{K \left(\tau_u \rho_u + 1\right)}}, \quad &\text{for ZF}. 
%\end{cases}
%\end{aligned}
%\end{equation}
\begin{equation} \label{eq:beta}
\beta = \left(\sqrt{\mathbb{E}{\left[{\mathrm{Tr}\left({\mathbf{U}}^\mathrm{H}{\mathbf{U}}\right)}\right]}}\right)^{-1}.
\end{equation}

\subsection{Received Signal at UE under NRC} \label{sec:NRC effect}
The multi-user received downlink signal vector is given by the second line of \eqref{eq:system_model}. Plugging the precoded symbol vector expression in \eqref{eq:precoded_symbols} into \eqref{eq:system_model}, the received signal for $k$'th user corresponding to the $k$'th element of $\mathbf{r}$ can be written as 
\begin{equation} \label{eq:received_signal}
{r_k} = \sqrt{\rho_d}{\beta}\mathbf{h}^\mathrm{T}_k{\mathbf{u}}_{k}s_k + \sqrt{\rho_d}{\beta}\sum\limits_{i = 1,i \ne k}^K {\mathbf{h}}^\mathrm{T}_k{\mathbf{u}}_{i}s_i  + z_{d,k},
\end{equation}
where $\mathbf{u}_k$ and $\mathbf{h}^\mathrm{T}_k$ denote the $k$'th column and row vectors of the precoder and effective downlink channel matrices, respectively. Notice that by denoting the $k$'th column of the uplink effective channel matrix as $\mathbf{g}_k$, the effective downlink channel towards the $k$'th user can be expressed as
\begin{equation} \label{eq:channel2k}
\mathbf{h}^\mathrm{T}_k = a_k \mathbf{g}_k^\mathrm{T} \mathbf{B}.% = a_k \left(\hat{\mathbf{h}}^\mathrm{T}_k + {\bm{\varepsilon}}^\mathrm{T}_k \right) \mathbf{B}.
\end{equation}

%Here, we assume that there are no downlink pilots transmitted towards UEs and UEs are using only the information of $\sqrt{\rho_d}\mathbb{E}\left[{\beta}\mathbf{h}_k{{\mathbf{u}}}_{k}\right]$ for the detection. Building on this and plugging \eqref{eq:channel2k} into \eqref{eq:received_signal}, the received signal can be re-written in the general form as
In general, conventional MIMO systems employ downlink pilots to acquire downlink CSI for detection purposes. However, in massive MIMO systems, as shown in \cite{yang_performance_2013, zhang_large-scale_2015, Jose_statistical_DL_1, Jose_statistical_DL_2, Larsson_no_DL_pilot}, it is generally assumed that UEs employ only the statistical properties of the beamformed channel, namely $\mathbb{E}\left[{\beta}\mathbf{h}^\mathrm{T}_k{{\mathbf{u}}}_{k}\right]$, as the downlink CSI to decode the received signal. This assumption is justified by the law of large numbers which implies that $\mathbf{h}^\mathrm{T}_k{{\mathbf{u}}}_{k} \rightarrow \mathbb{E}\left[\mathbf{h}^\mathrm{T}_k{{\mathbf{u}}}_{k}\right]$, commonly known as the channel hardening concept \cite{marzetta_ch_hardening,HienNgo_massMIMO_DLpilot}. Utilizing such approach in acquiring downlink CSI in UEs eliminates the need for sending downlink pilots which directly reduces downlink overhead. Building on this and plugging \eqref{eq:channel2k} into \eqref{eq:received_signal}, the received signal under NRC can be re-written in a general form as
\begin{equation} \label{eq:received_signal_k}
{r_k} = \sqrt{\rho_d}\mathbb{E}\left[{\beta}\mathbf{h}^\mathrm{T}_k{{\mathbf{u}}}_{k}\right] s_k + z_{\mathrm{SI}, k} + z_{\mathrm{IUI},k}+z_{d,k},
\end{equation}
where the self-interference (SI), $z_{\mathrm{SI}, k}$, and inter-user-interference (IUI), $z_{\mathrm{IUI},k}$, are given by %expressions in \eqref{eq:SI-IUI} at the bottom of the next page.
\begin{equation}\label{eq:SI_IUI}
\begin{aligned}
z_{\mathrm{SI},k} &= \sqrt{\rho_d}{\beta}\left(a_k {\mathbf{g}}^\mathrm{T}_k \mathbf{B}\mathbf{u}_{k}-\mathbb{E}\left[\mathbf{h}^\mathrm{T}_k\mathbf{u}_{k}\right]\right)s_k \\
z_{\mathrm{IUI},k} &= \sqrt{\rho_d}{\beta}\sum\limits_{i = 1,i \ne k}^K a_k{\mathbf{g}}^\mathrm{T}_k \mathbf{B}\mathbf{u}_{i} s_i.
\end{aligned}
\end{equation}
Based on \eqref{eq:SI_IUI}, it can be clearly observed that the NRC-blind precoder $\mathbf{u}$ which is constructed based on the estimated uplink effective channel ${\hat{\mathbf{G}}}$, through $\hat{\mathbf{H}}=\hat{\mathbf{G}}^\mathrm{T}$, cannot take into account the NRC effects from $a_k$ and $\mathbf{B}$, which results into increased interference levels and thus reduced downlink spectral efficiency. This is illustrated through an elementary system spectral efficiency evaluation in \figurename{~\ref{fig:Seff_vs_SNR_best_naiveBF}}, with the detailed evaluation assumptions being described in Section \ref{sec:simulations}. It can be noticed that in particular in the ZF precoder case, NRC-blind precoding results into substantial performance degradation, hence strongly motivating to develop efficient NRC estimation and mitigation techniques.

%The effects of the NRC on MRT and ZF-based precoded downlink transmission can thus be summarized as follows \cite{our_submitted_analysis_journal}:
%\begin{enumerate}
%\item MRT-based transmission is known to have IUI even under perfect CSI and without NRC. However, under NRC, the IUI term becomes a function of the NRC variables in addition to channel and estimation noise variables. Moreover, there is an additional SI-term compared to SI-free reception without NRC.
%
%\item ZF-based transmission is known to be IUI-free with perfect CSI for reciprocal channels. However, under NRC, there is IUI even with perfect CSI. In the most general case with imperfect CSI and with NRC, the IUI is a function of channel, estimation noise and NRC variables. Moreover, there is additional SI term compared to SI-free reception without NRC.
%\end{enumerate}

\subsection{NRC-Aware Downlink Precoding Principle} \label{sec:mitigation}
As shown in Section \ref{sec:NRC effect} above, if MRT and/or ZF precoders are applied naively without accounting for NRC, there are additional SI and IUI terms that can substantially degrade the quality of the received signal at the UE side. Here, we introduce a novel NRC mitigation approach, called NRC-aware precoding, which seeks to cancel out the effects of NRC by properly modifying the precoder.

Assuming that the BS has already estimates of the NRC matrices ${\mathbf{A}}$ and ${\mathbf{B}}$, denoted by $\hat{\mathbf{A}}$ and $\hat{\mathbf{B}}$, the NRC-aware precoding approach transforms the basic linear precoders given in \eqref{eq:MRT_ZF} as
\begin{equation} \label{eq:NRC_mitigator}
\mathbf{U}_{nrc} = \hat{\mathbf{B}}^{-1}\mathbf{U}\hat{\mathbf{A}}^{-1}.
\end{equation}
%\begin{equation} \label{eq:NRC_mitigator}
%\begin{aligned} 
%\mathbf{U}_{nrc} = \begin{cases} \left(\mathbf{B}^*\right)^{-1}\mathbf{U}\left(\mathbf{A}^*\right)^{-1},\quad &\text{for MRT}\\
%\left(\mathbf{B}^\mathrm{T}\right)^{-1}\mathbf{U}\left(\mathbf{A}^\mathrm{T}\right)^{-1},\quad &\text{for ZF}.
%\end{cases}
%\end{aligned}
%\end{equation}
Note that, in the special case where the NRC estimation method is capable of estimating the BS side NRC only, \eqref{eq:NRC_mitigator} reduces to $\mathbf{U}_{nrc} = \hat{\mathbf{B}}^{-1}\mathbf{U}$.

\begin{figure}[!t]
	\centering
	\includegraphics[width=0.48\textwidth]{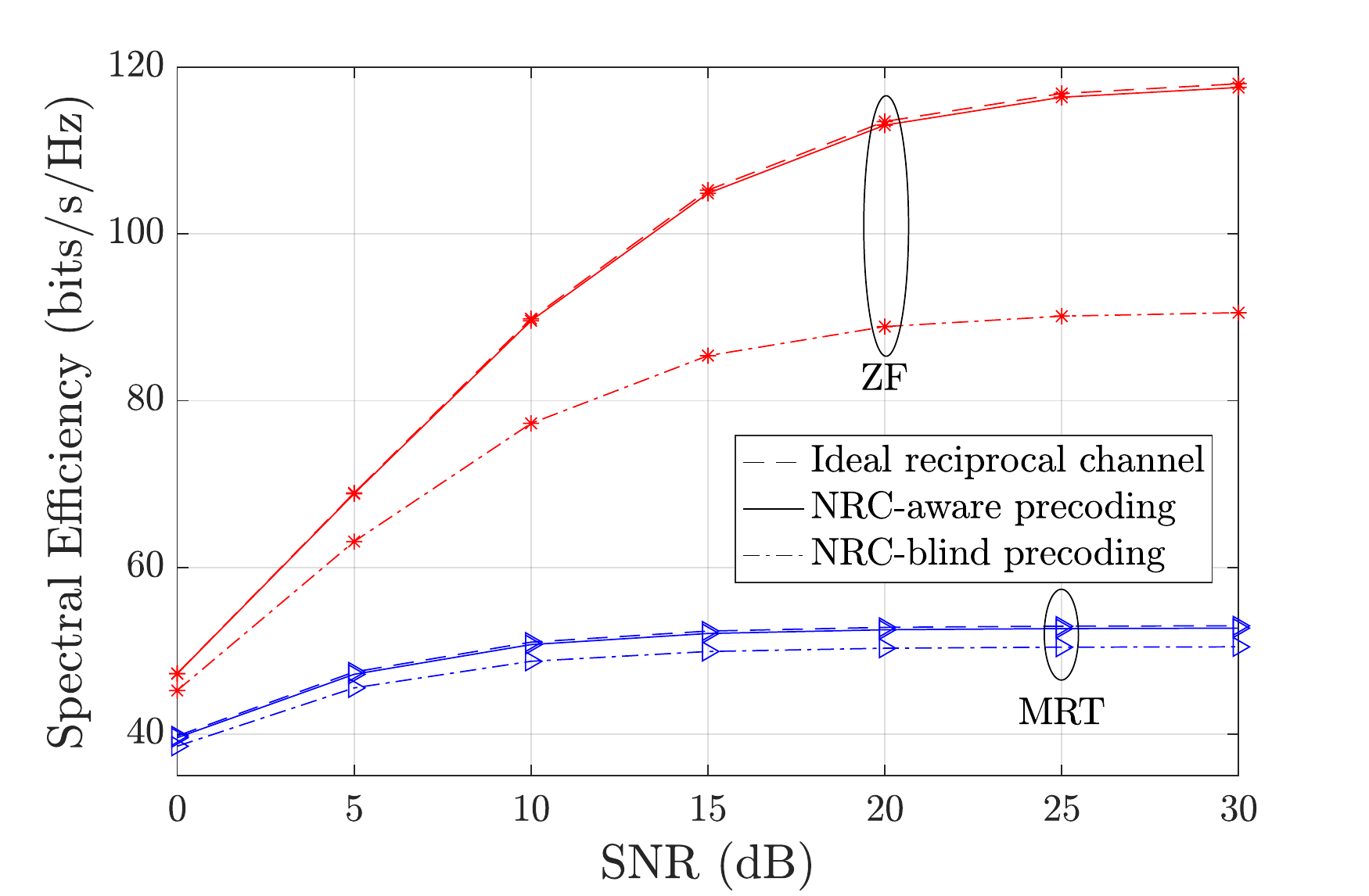}
	\caption{Spectral efficiency vs. downlink SNR ($\rho_d$) for $N = 100$, $K = 20$, $\tau_u = K$, $\rho_u = 0$ dB, $T = 250$.}
	\label{fig:Seff_vs_SNR_best_naiveBF}
\end{figure}
The system spectral efficiency performance with the NRC-aware precoder, assuming ideal NRC estimates, is shown in \figurename{~\ref{fig:Seff_vs_SNR_best_naiveBF}}. As can be observed, the NRC-aware precoder achieves the ideal system performance, i.e., the performance with fully reciprocal channels. 
%The spectral efficiency degradation due to SI and IUI are illustrated in \figurename{~\ref{fig:Seff_vs_SNR_best_naiveBF}} when MRT and ZF-based precoders are applied naively, without addressing the estimation/mitigation of NRC. These results point out the need for NRC-aware precoders, especially in case of ZF precoding. As illustrated, the NRC mitigation technique is defined in a way that it achieves approximately the identical benchmark performances for reciprocal channels, which can be obtained from \eqref{eq:received_signal_k} by settings $\mathbf{A}=\mathbf{B}=\mathbf{I}$ and $\mathbf{A}^{\prime}=\mathbf{B}^{\prime}=\mathbf{0}$, when the NRC matrices estimation is ideal.
The evaluation setup and details of spectral efficiency calculations will be described in Section \ref{sec:simulations}.

\section{Proposed Estimation of NRC Matrices} \label{sec:estimation_mitigation}
%\subsection{Acquisition of UE and BS-side NRC Matrices $\hat{\mathbf{A}}$ and $\hat{\mathbf{B}}$ at BS}
The NRC mitigation method, i.e., the NRC-aware precoder described in Section \ref{sec:mitigation} requires the knowledge of the matrices ${\mathbf{A}}$ and ${\mathbf{B}}$ at the BS. The information about these matrices is not readily available, hence calling for efficient estimation approaches. Thus, in this section, we will propose a novel iterative OTA estimation framework for acquiring accurate estimates of  ${\mathbf{A}}$ and ${\mathbf{B}}$, based on a novel pilot signaling concept between the BS and UEs.

In general, the NRC variances $\sigma_\mathrm{F}^2$, $\sigma_\mathrm{L}^2$, $\sigma_\mathrm{M}^2$, and the corresponding realizations of the elements of $\mathbf{A}$ and $\mathbf{B}$ depend on hardware characteristics and operating conditions, e.g., temperature, which vary slowly in time. Thus, the NRC characteristics and the corresponding realizations of $\mathbf{A}$ and $\mathbf{B}$ can be assumed to stay constant over many propagation channel coherence intervals \cite{guillaud_practical_2005}. Therefore, it is sufficient to perform the NRC estimation very infrequently, e.g, once in every $10$ minutes or once a day \cite{shepard_argos:_2012,Rogalin_rel_calib}, which makes the estimation overhead negligible, when compared to signaling and pilot overhead that commonly rises from channel estimation procedures.

%As mentioned in Section \ref{sec:introduction}, there are mainly two types of NRC mitigation methods, namely self-calibration and over-the-air. Self-calibration methods, in general, focus on mitigating BS NRC while assuming UE NRC does not have major impact on the system performance \cite{Rogalin_rel_calib, vieira_reciprocity_2014, wei_mutual_2016, wei_tdd_2015}. However, with the absence of downlink pilots in the system, UE NRC becomes one of the main causes of performance degradation \cite{our_submitted_analysis_journal}. Therefore, in this work, we prefer an OTA-based method where the only drawback is its overhead which, as discussed earlier, is negligible.

\subsection{Proposed Pilot Signaling}
In order to estimate the matrices ${\mathbf{A}}$ and ${\mathbf{B}}$, we propose the following round-trip pilot signaling approach:
\begin{enumerate}
	\item BS transmits an $N\times N$ orthonormal pilot matrix $\mathbf{X}_{nrc}$.
	\item Upon reception, without decoding, UEs send back the conjugated versions of the received signals.
\end{enumerate}

Based on the above scheme, the received multi-user signal matrix at UE side can be written as 
\begin{equation} \label{eq:NRC_pilot_down}
\mathbf{R}_{nrc}=\sqrt{\tilde{\rho}_{d}}\mathbf{H}\mathbf{X}_{nrc}+\tilde{\mathbf{Z}}_d,
\end{equation}
where $\tilde{\rho}_{d}$ is the downlink SNR and $\tilde{\mathbf{Z}}_d$ is the $K \times N$ multi-user receiver noise matrix with i.i.d. $\mathcal{CN}\left(0,1\right)$ entries. The tilde sign is used in above and what follows to distinguish these variables between the actual data transmission and pilot signaling phases. Then, the corresponding received signal at BS with the UEs sending back the conjugated version of \eqref{eq:NRC_pilot_down} reads
\begin{equation} \label{eq:NRC_pilot_up}
\begin{aligned} 
\mathbf{Y}_{nrc}&=\sqrt{\tilde{\rho}_{u}}\mathbf{G}\mathbf{R}^{*}_{nrc}+\tilde{\mathbf{Z}}_u \\
&=\sqrt{\tilde{\rho}_{u}}\sqrt{\tilde{\rho}_{d}}\mathbf{G}\mathbf{H}^{*}\mathbf{X}^{*}_{nrc}+\tilde{\mathbf{Z}}_{tot},
\end{aligned}
\end{equation}
where $\tilde{\rho}_{u}$ is the uplink SNR and $\tilde{\mathbf{Z}}_u$ is the $N \times N$ BS receiver noise matrix with i.i.d. $\mathcal{CN}\left(0,1\right)$ entries. The total effective noise matrix seen at BS is denoted as $\tilde{\mathbf{Z}}_{tot}=\sqrt{\tilde{\rho}_{u}}\mathbf{G}\tilde{\mathbf{Z}}^{*}_d+\tilde{\mathbf{Z}}_u$.

\begin{figure*}[t]
	\centering
	\includegraphics[width=0.7\textwidth]{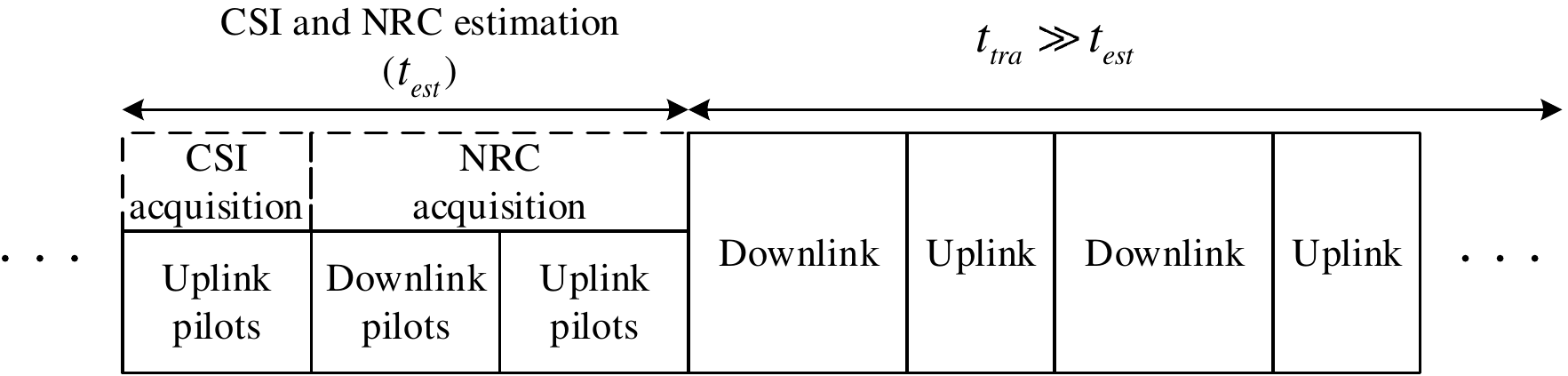}
	\caption{Assumed radio frame or sub-frame structure incorporating CSI and NRC estimation as well as actual data transmission phases.}
	\label{fig:calib_timeline}
\end{figure*}
In above, the duration of the described overall NRC-related pilot signaling is $2N$ symbols where the uplink and downlink channels are assumed to be fixed. The coherence time of the physical channels is typically in the order of several hundred symbol intervals, determined mostly by the mobility of the UEs and the system center-frequency. Hence, we assume a %low to low-medium mobility 
scenario where the coherence time is at least $2N+K$ symbols, taking into account both NRC-related pilot signaling and uplink channel estimation. As mentioned in the previous section, matrices $\mathbf{A}$ and $\mathbf{B}$ are expected to change very slowly compared to channel coherence time and hence it is assumed that their values are fixed during the above pilot signaling. \figurename{~\ref{fig:calib_timeline}} illustrates the overall assumed radio frame or sub-frame structure of the considered massive MIMO TDD system including the proposed NRC estimation phase.

\subsection{Overall Estimation Framework} \label{sec:iterative}
As the initial step in estimating $\mathbf{A}$ and $\mathbf{B}$, the BS processes the received signal $\mathbf{Y}_{nrc}$ in \eqref{eq:NRC_pilot_up} as $\mathbf{Q}=\mathbf{Y}^{*}_{nrc}\mathbf{X}^\mathrm{H}_{nrc}$. Since the pilot matrix $\mathbf{X}_{nrc}$ has the property $\mathbf{X}^\mathrm{H}_{nrc}\mathbf{X}_{nrc}=\mathbf{I}_N$, the processed signal can be expressed as
\begin{equation} \label{eq:NRCprocessed}
\mathbf{Q}=\sqrt{\tilde{\rho}_{u}}\sqrt{\tilde{\rho}_{d}}\mathbf{G}^{*}\mathbf{A} \mathbf{G}^\mathrm{T} \mathbf{B}+\mathbf{V},
\end{equation}        
where the processed noise matrix is given by $\mathbf{V}=\tilde{\mathbf{Z}}^{*}_{tot}\mathbf{X}^\mathrm{H}_{nrc}$.

Now the target is to estimate $\mathbf{A}$ and $\mathbf{B}$ from \eqref{eq:NRCprocessed} assuming that the BS has the uplink channel estimate $\hat{\mathbf{G}}$. In this respect, denoting the estimates at $m$'th iteration as $\hat{\mathbf{A}}(m)$ and $\hat{\mathbf{B}}(m)$, we propose the following iterative estimation framework:
\begin{enumerate}
	\item Initialize, $\hat{\mathbf{A}}(0)=\mathbf{I}_K$, obtain the estimate $\hat{\mathbf{B}}(1)$.  
	\item Substitute $\hat{\mathbf{B}}(1)$ for ${\mathbf{B}}$ in \eqref{eq:NRCprocessed} and obtain estimate $\hat{\mathbf{A}}(1)$.
	\item Successively refine the estimates $\hat{\mathbf{A}}$ and $\hat{\mathbf{B}}$ by fixing the current value of one and solving for the other from \eqref{eq:NRCprocessed}.       
\end{enumerate}

In above, $\mathbf{I}_K$ is used for initialization since the deviation matrix $\mathbf{A}^{\prime}$ in $\mathbf{A}=\mathbf{I}_K+\mathbf{A}^{\prime}$ is in practice small. Notice that the processed received signal in \eqref{eq:NRCprocessed} and the corresponding UL channel estimate are available at multiple parallel sub-carriers in an OFDM/OFDMA based radio system. Hence, the above iterative estimation scheme can be carried out in a per subcarrier manner as well. Furthermore, as mentioned in \cite{petermann_multi-user_2013}, transceivers' baseband-to-baseband behavior can be modeled by allpass-like transfer functions, therefore it is reasonable to assume that the NRC matrices $\mathbf{A}$ and $\mathbf{B}$ are largely similar over a set of adjacent subcarriers $C_{sc}$ where typically $C_{sc} \leq 10$, whereas $\mathbf{G}$ is subject to variations depending on the frequency selectivity of the propagation channels. Based on these assumptions, the estimates $\hat{\mathbf{A}}$ and $\hat{\mathbf{B}}$ can be obtained by averaging the per subcarrier estimates over $C_{sc}$ neighboring subcarriers, i.e., 
\begin{equation}
\begin{aligned}
\hat{\mathbf{A}} = \frac{1}{C_{sc}} \sum_{l=1}^{C_{sc}} \hat{\mathbf{A}}_l \\
\hat{\mathbf{B}} = \frac{1}{C_{sc}} \sum_{l=1}^{C_{sc}} \hat{\mathbf{B}}_l,
\end{aligned}
\end{equation}
where $l$ denotes the subcarrier index. Next we will present the actual proposed methods to obtain the estimates for $\mathbf{A}$ and $\mathbf{B}$. To simplify the notation, we will drop the subcarrier index $l$.

\subsection{Proposed Estimation of $\mathbf{B}$} \label{sec:B estimation}
As described earlier, $\hat{\mathbf{B}}$ is iteratively refined using the current estimate of $\mathbf{A}$. The proposed estimator builds on solving the matrix equation in \eqref{eq:NRCprocessed} based on minimizing the Frobenius norm criterion. With this setting, the refined estimate of $\mathbf{B}$ can be formulated as
\begin{equation} \label{eq:Best}
\hat{\mathbf{B}}(m+1)=\underset{\mathbf{B}}{\text{argmin}} \left|\left| \mathbf{Q}-\sqrt{\tilde{\rho}_{u}}\sqrt{\tilde{\rho}_{d}}\hat{\mathbf{G}}^*\hat{\mathbf{A}}(m) \hat{\mathbf{G}}^\mathrm{T} {\mathbf{B}}\right|\right|^2_F,
\end{equation}
where the subscript $F$ denotes the Frobenius norm.

Next, by denoting ${\mathbf{T}}(m)=\sqrt{\tilde{\rho}_{u}}\sqrt{\tilde{\rho}_{d}}\hat{\mathbf{G}}^*\hat{\mathbf{A}}(m) \hat{\mathbf{G}}^\mathrm{T}$, we have the following identity
\begin{equation} \label{eq:Bidentity}
\left|\left| \mathbf{Q}- {\mathbf{T}}(m){\mathbf{B}}\right|\right|^2_F =\sum_{j=1}^{K} \left|\left| \mathbf{q}_j- {\mathbf{T}}(m){\mathbf{b}_j} \right|\right|^2,  
\end{equation}
where $\mathbf{q}_j$ and ${\mathbf{b}_j}$ denote the $j$'th column of $\mathbf{Q}$ and $\mathbf{B}$, respectively. Since the $j$'th term in the sum depends only on ${\mathbf{b}_j}$, minimizing the total sum is equivalent to separately minimizing each term $\left|\left| \mathbf{q}_j- {\mathbf{T}}(m){\mathbf{b}_j}\right|\right|^2$. Thus, the estimation of matrix $\mathbf{B}$ is eventually simplified to estimation of each column of $\mathbf{B}$, independently.  

\begin{figure}[!t]
	\centering
	\includegraphics[width=0.48\textwidth]{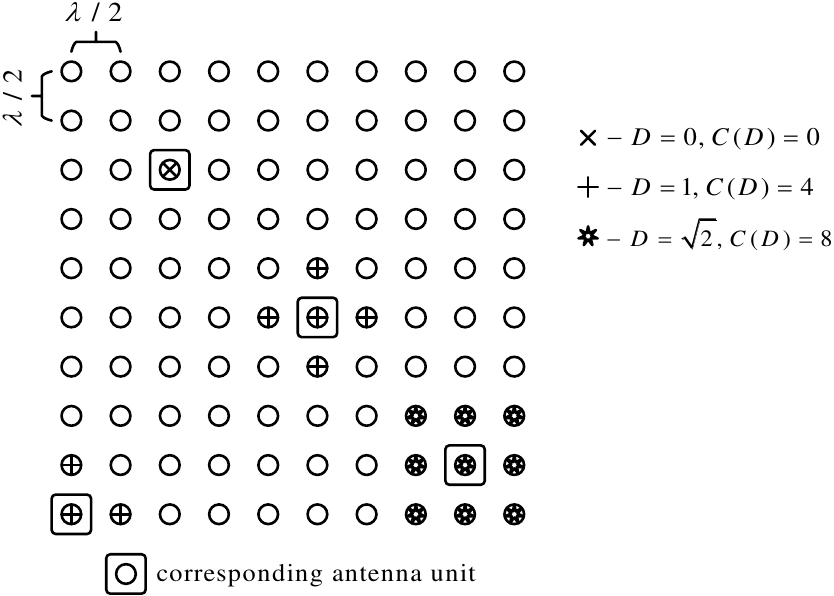}
	\caption{Illustration of sparsity threshold $D$ on $10 \times 10$ rectangular BS antenna grid with $\lambda/2$ antenna spacing.}
	\label{fig:D}
\end{figure}
As mentioned earlier, the BS NRC matrix incorporates both the frequency-responses and mutual coupling at the BS side. The power of mutual coupling between two different antenna units is related to their physical distance, thus the off-diagonal elements in $\mathbf{B}$ become smaller as the distance between the two corresponding antenna units increases. Here, in estimating the BS NRC matrix $\mathbf{B}$, we treat those off-diagonal entries which are corresponding to two antennas with a distance larger than a pre-defined threshold, called sparsity threshold $D$, as zeros, yielding a sparse matrix structure for $\hat{\mathbf{B}}$. We also define the maximum number of coupled neighboring antenna elements as $C(D)$. In \figurename{~\ref{fig:D}} an example $10 \times 10$ rectangular antenna layout with $\lambda/2$ antenna spacing between the neighboring elements is shown with $3$ different values of $D$, namely $D = 0$, $1$ and $\sqrt{2}$, measured as multiples of $\lambda/2$. When $D = 0$, it is assumed that there is no mutual coupling and $C(D)=0$, whereas for $D = 1$ and $D = \sqrt{2}$, the central antenna elements are coupled with at most $C(D)=4$ and $C(D)=8$ closest neighboring antenna elements. Note that, the antenna elements close to the edges of the grid are coupled with less number of antenna units. This is illustrated in \figurename{~\ref{fig:D}} where for $D = 1$, the bottom left antenna element is assumed to be coupled with only $2 \leq C(1) = 4$ antennas.

The following column-wise estimator will build on the assumption that ${\mathbf{B}}$ has a sparse structure and the number of non-zero row entries in any column $j$, denoted as $R_j$, satisfies
\begin{equation}\label{eq:rank}
R_j \le R\left(D\right),
\end{equation} 
where $R(D) = C(D) + 1$.
It is further assumed that the index of non-zero entries of $\mathbf{b}_j$ are known, which is directly determined by the antenna system architecture and geometry, and the assumed pre-defined coupling threshold discussed above. With these assumptions, we define a reduced vector of dimension $R_j \times 1$, ${\mathbf{b}^{red}_j}$, that contains the non-zero entries of $\mathbf{b}_j$. If the $j$'th row is kept when constructing ${\mathbf{b}^{red}_j}$, then similarly, the $j$'th column is kept to construct ${\mathbf{T}}^{red}_j(m)$. Based on these, we can formulate the estimation of columns of $\hat{\mathbf{B}}(m+1)$ through a reduced system of equations as
\begin{equation} \label{eq:Breduced}
\hat{\mathbf{b}}^{red}_j(m+1)=\underset{\mathbf{b}^{red}_j}{\text{argmin}} \left|\left| \mathbf{q}_j- {\mathbf{T}}^{red}_j(m){\mathbf{b}^{red}_j} \right|\right|^2.
\end{equation}

The solution to \eqref{eq:Breduced} is then given by
\begin{equation} \label{eq:BLS}
\hat{\mathbf{b}}^{red}_j(m+1)=\left( {\mathbf{T}}^{red}_j(m)\right)^{\dagger}\mathbf{q}_j.
\end{equation}
Once the $\hat{\mathbf{b}}^{red}_j(m+1)$ is solved from \eqref{eq:BLS}, then $\hat{\mathbf{b}}_j(m+1)$ can be obtained straightforwardly by appending zeros to the appropriate rows.

Note that, when $\mathbf{A} \approx \mathbf{I}_K$, we also have $\mathbf{G}^{*}\mathbf{A} \mathbf{G}^\mathrm{T} \approx \mathbf{G}^{*}\mathbf{G}^\mathrm{T}$, where the matrix $\mathbf{G}^{*}\mathbf{G}^\mathrm{T}=\left(\mathbf{G}\mathbf{G}^\mathrm{H}\right)^{*}$ is positive semi-definite matrix and of rank $K$ if $\mathbf{G}$ is of rank $K$. The obtained $\hat{\mathbf{b}}^{red}_j$ and the corresponding minimum expression from \eqref{eq:Breduced} depend on the corresponding values of $K$ and $D$. The column space of ${\mathbf{T}}^{red}_j(m)$ has higher dimensionality for larger $K$. Thus, when $D$ is fixed, for larger $K$ one can solve for $\hat{\mathbf{b}}^{red}_j$ from \eqref{eq:BLS} which yields smaller values of $\left|\left| \mathbf{q}_j- {\mathbf{T}}^{red}_j(m){\mathbf{b}^{red}_j} \right|\right|^2$.

\subsection{Proposed Estimation of $\mathbf{A}$} 
Next, given $\hat{\mathbf{B}}(m)$ from \eqref{eq:Best}, the (refined) estimate of $\mathbf{A}$ can be formulated based on minimizing the Frobenius norm criterion as
\begin{equation} \label{eq:Aestimation}
\hat{\mathbf{A}}(m)=\underset{\mathbf{A}}{\text{argmin}} \left|\left| \mathbf{Q}-\sqrt{\tilde{\rho}_{u}}\sqrt{\tilde{\rho}_{d}}\hat{\mathbf{G}}^*{\mathbf{A}} \hat{\mathbf{G}}^\mathrm{T} \hat{\mathbf{B}}(m)\right|\right|^2_F.
\end{equation}
For diagonal $\mathbf{A}$, the solution to \eqref{eq:Aestimation} can be obtained as
\begin{equation}
\hat{\mathbf{A}}(m)= \mathrm{diag}\left(\hat{\boldsymbol{\xi}}\right).
\end{equation} 
where $\hat{\boldsymbol{\xi}}=\left[\mathbf{I}_K, i\mathbf{I}_K\right]\hat{\boldsymbol{\psi}}$ and the vector $\hat{\boldsymbol{\psi}}$ is given as
\begin{equation} \label{eq:UE NRC solution}
\hat{\boldsymbol{\psi}}=\left(\sum_{j=1}^{N}  \bar{\mathbf{W}}_j^\mathrm{T}\bar{\mathbf{W}}_j\right)^{-1} \sum_{j=1}^{N}  \bar{\mathbf{W}}_j^\mathrm{T}\bar{\mathbf{q}}_j.
\end{equation}
In above, $\bar{\mathbf{q}}_j=\left[\Re\{\mathbf{q}^\mathrm{T}_j\},\Im\{\mathbf{q}^\mathrm{T}_j\}\right]^\mathrm{T}$, and defining the $N \times N$ matrix $\mathbf{W}_j=\sqrt{\tilde{\rho}_{u}}\sqrt{\tilde{\rho}_{d}}\hat{\mathbf{G}}^*\mathrm{diag}\left(\left[\hat{\mathbf{G}}^\mathrm{T}\hat{\mathbf{B}}(m)\right]_j\right)$, with $\left[\hat{\mathbf{G}}^\mathrm{T}\hat{\mathbf{B}}(m)\right]_j$ being the $j$'th column of $\hat{\mathbf{G}}^\mathrm{T}\hat{\mathbf{B}}(m)$, $\bar{\mathbf{W}}_j$ is given as 
\begin{equation}
\bar{\mathbf{W}}_j=\begin{bmatrix}
\Re\{\mathbf{W}_j\}, &-\Im\{\mathbf{W}_j\} \\
\Im\{\mathbf{W}_j\}, &\Re\{\mathbf{W}_j\}
\end{bmatrix}.
\end{equation}

\textit{Proof:} See Appendix.

\section{Numerical Evaluations and Analysis} \label{sec:simulations} 
\subsection{Basic Evaluation Settings and Performance Measures}
In this section, by using extensive computer simulations, we evaluate the performance of the proposed NRC estimation and mitigation scheme. We also compare its performance to the performance of two other existing schemes in literature, namely the direct-path based least squares (LS) known as ``Argos'' \cite{shepard_argos:_2012} and the generalized neighbor LS \cite{vieira_reciprocity_2014}. The latter is the optimized version of the generalized LS method presented in \cite{Rogalin_rel_calib} and is shown in \cite{vieira_reciprocity_2014} to have the best performance amongst several existing NRC estimation methods. Both LS based methods estimate the BS NRC by the means of mutual coupling between BS antennas, while they depend on the downlink pilots to compensate the NRC in the UE side.

We consider the DL spectral efficiency as the key performance metric, which is defined as
\begin{equation} \label{eq:spectralEfficiency}
\eta_s = K \left(1 - \frac{\tau_u + \tau_d}{T}\right) \mathbb{E}\left[\log_2\left(1 + \mathrm{SINR}\right)\right],
\end{equation}
where the expectation is over different NRC realizations and channel coherence intervals. The length of downlink pilots in symbols is denoted by $\tau_d$ and $T$ is the total number of symbols in a channel coherence interval. $\mathrm{SINR}$ is the instantaneous signal to interference and noise ratio (SINR), which can be written, based on \eqref{eq:received_signal}, as
\begin{equation}\label{eq:instSINR}
\mathrm{SINR} = \frac{\left|\sqrt{\rho_d}\hat{\alpha}_k s_k\right|^2}{\left|r_k - \sqrt{\rho_d}\hat{\alpha}_k s_k\right|^2},
\end{equation}
where $\hat{\alpha}_k$ is the scaling of the useful signal term available at the receiver of the $k$'th UE. In the context of the proposed NRC estimation and mitigation method, no DL CSI pilots are used. Hence, for the proposed estimation method, $\hat{\alpha}_k = \mathbb{E}\left[{\beta}\mathbf{h}^\mathrm{T}_k{{\mathbf{u}}}_{k}\right]$. On the contrary, the other two estimation methods utilizes $\tau_d \geq K$ downlink pilots for DL CSI acquisition as described in \cite{HienNgo_massMIMO_DLpilot}. %In practice, there is a residual error between the estimated and true scaling factors due to finite downlink SNR.

The other relevant performance metric is the normalized mean squared error (MSE) for NRC estimation which is defined as
\begin{equation}
\begin{aligned}
\delta_e^2 = \begin{cases}\dfrac{\left|\left| \mathbf{B} - \hat{\mathbf{B}} \right|\right|^2_F}{\left|\left| \mathbf{B} \right|\right|^2_F},\quad &\text{for BS side}\\
\dfrac{\left|\left| \mathrm{diag}\left(\mathbf{A}\right) - \mathrm{diag}\left(\hat{\mathbf{A}}\right) \right|\right|^2_F}{\left|\left| \mathrm{diag}\left(\mathbf{A}\right) \right|\right|^2_F},\quad &\text{for UE side}.
\end{cases}
\end{aligned}
\end{equation}
%where $\hat{\alpha}_k$ is the estimated version of the downlink channel ${\alpha}_k = {\beta}\mathbf{h}_k{{\mathbf{u}}}_{k}$. To evaluate the performance of the proposed NRC estimation method where there is no downlink pilots in the system ($\tau_d=0$), $k$'th UE employs statistical properties of the channel as $\hat{\alpha}_k = \mathbb{E}\left[{\beta}\mathbf{h}_k{{\mathbf{u}}}_{k}\right]$ to decode the received signal, whereas the downlink channel estimation method presented in \cite{HienNgo_massMIMO_DLpilot} is used for the other two NRC estimation methods which requires the availability of downlink pilots with $\tau_d \geq K$.

As a baseline simulation scenario, we consider a BS which is equipped with $N = 100$ infinitely thin $\lambda/2$ dipole antennas in a $10 \times 10$ square layout with $\lambda/2$ spacing as illustrated in \figurename{~\ref{fig:D}}. The input and the mutual impedances are computed based on \cite{dipole_ant_book} for the assumed carrier-frequency of $f_c = 3.5$ GHz. The impedances are assumed to be frequency-independent, as the modulated signal RF bandwidth is much smaller than the carrier frequency. The uplink channel matrix $\mathbf{G}$ is assumed to have i.i.d. $\mathcal{CN}\left(0,1\right)$ elements. The BS serves $K = 20$ single-antenna UEs simultaneously on the same time-frequency resource through either ZF or MRT precoding. We assume a %low-medium mobility 
scenario where each coherence interval contains $T = 250$ OFDM symbols. The number of uplink pilots sent by each UE in each coherence interval is equal to the number of scheduled UEs, $\tau_u = K$, and the uplink SNR in this phase is assumed to be $\rho_u = 0$ dB. In the scenarios where UEs rely on downlink pilots for decoding purposes, i.e., direct-path based LS and generalized neighbor LS methods, the number of downlink pilots in each coherence interval is set to be $\tau_d = K$ \cite{HienNgo_massMIMO_DLpilot}, and their SNR is equal to the downlink SNR in data transmission phase which is assumed to be $\rho_d = 20$ dB. The SNR of the coupling channel between two neighboring antennas is set to be $80$ dB for the two mentioned NRC mitigation methods \cite{vieira_reciprocity_2014}. The uplink and downlink SNRs for the pilot signaling in the proposed NRC estimation framework are set to be $\tilde{\rho}_{u} = 0$ dB and $\tilde{\rho}_{d} = 10$ dB, respectively. In the proposed method, the estimated NRC matrices are averaged over $10$ neighboring subcarriers, $C_{sc} = 10$, over which the NRC realizations are assumed to be constant. Finally, the variances of transceivers frequency-responses in both BS and UE side are assumed to be $-20$ dB, i.e., $\sigma_\mathrm{L}^2 = \sigma_\mathrm{F}^2 = -20$ dB. These are the baseline simulation settings, while some of the parameter values are also varied in the evaluations.

\subsection{Numerical Results}
\subsubsection{Effect of Sparsity Distance Threshold $D$}
%\begin{figure}[!t]
%	\centering
%	\subfloat[]{\includegraphics[width=0.48\textwidth]{MC_power_different_D_K_10_MSE.pdf}%
%		\label{subfig:MC_K_10_MSE}}
%	\hfil
%	\subfloat[]{\includegraphics[width=0.48\textwidth]{MC_power_different_D_MSE.pdf}%
%		\label{subfig:MC_K_20_MSE}}
%	\caption{NRC estimation normalized MSE vs. input reflection coefficients variance ($\sigma_\mathrm{M}^2$) for $N = 100$, \protect\subref{subfig:MC_K_10_MSE} $K = 10$ and \protect\subref{subfig:MC_K_20_MSE} $K = 20$, $\tau_u = K$, $\rho_u = 0$ dB, $T = 250$.}
%	\label{fig:MC_power_different_D_MSE}
%\end{figure}
%\begin{figure}[!t]
%	\centering
%	\subfloat[]{\includegraphics[width=0.48\textwidth]{MC_power_different_D_K_10.pdf}%
%		\label{subfig:MC_K_10}}
%	\hfil
%	\subfloat[]{\includegraphics[width=0.48\textwidth]{MC_power_different_D.pdf}%
%		\label{subfig:MC_K_20}}
%	\caption{Spectral efficiency vs. input reflection coefficients variance ($\sigma_\mathrm{M}^2$) for $N = 100$, \protect\subref{subfig:MC_K_10} $K = 10$ and \protect\subref{subfig:MC_K_20} $K = 20$, $\tau_u = K$, $\rho_u = 0$ dB, $T = 250$.}
%	\label{fig:MC_power_different_D}
%\end{figure}
\begin{figure}[!t]
	\centering
	\subfloat[]{\includegraphics[width=0.48\textwidth]{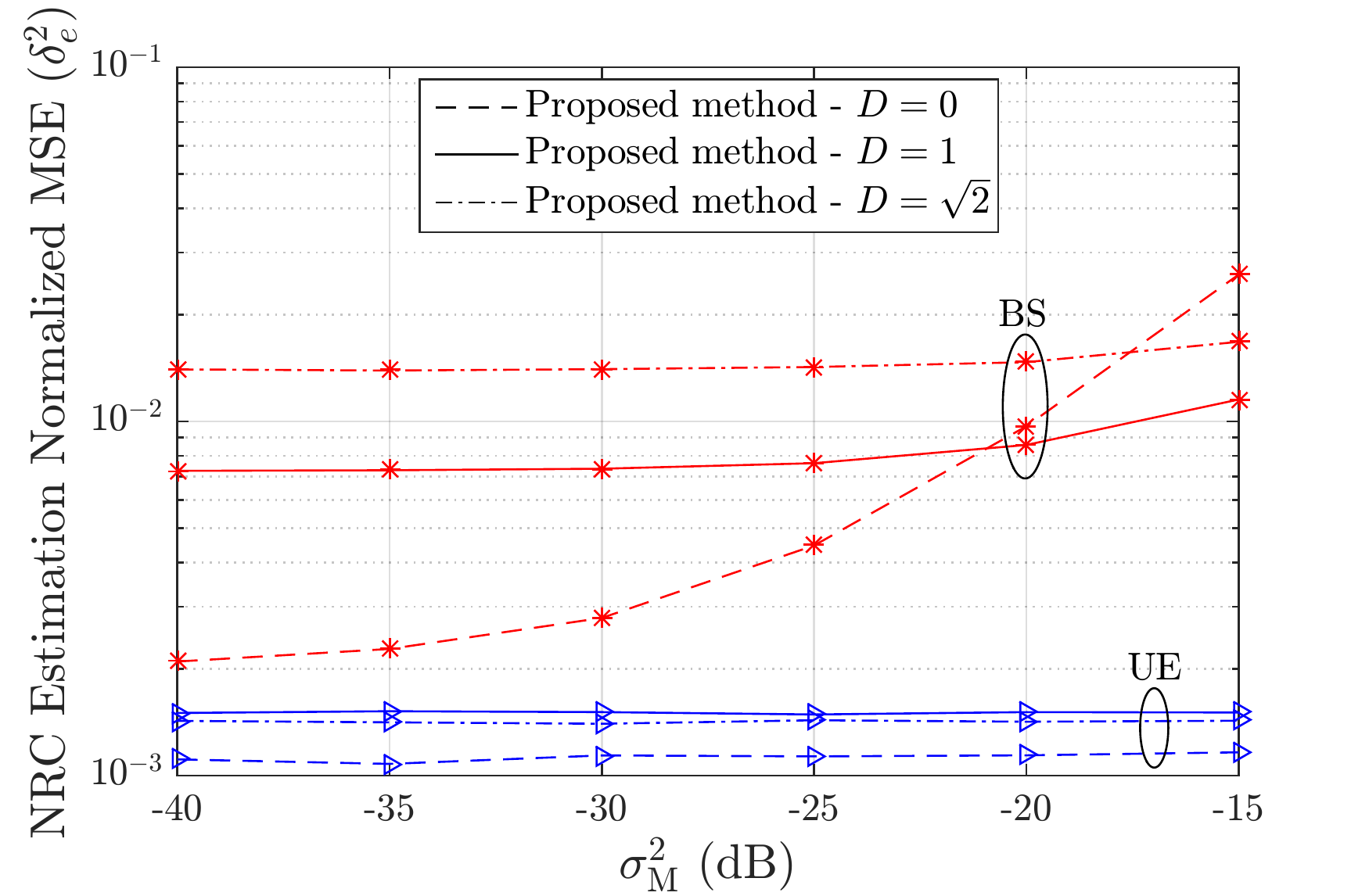}%
		\label{subfig:MC_K_20_MSE}}
	\hfil
	\subfloat[]{\includegraphics[width=0.48\textwidth]{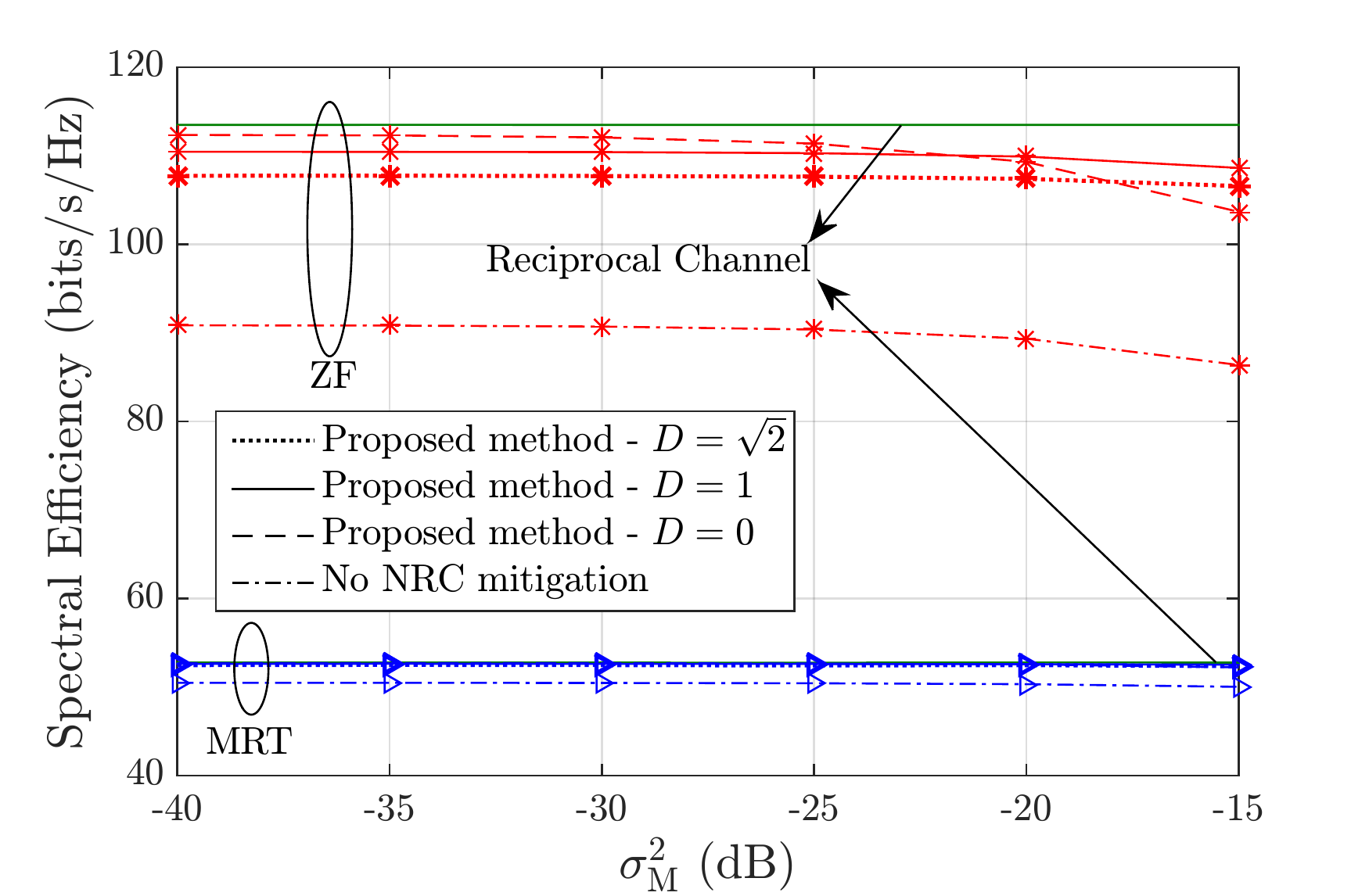}%
		\label{subfig:MC_K_20}}
	\caption{\protect\subref{subfig:MC_K_20_MSE} NRC estimation normalized MSE and \protect\subref{subfig:MC_K_20} system spectral efficiency vs. input reflection coefficients variance ($\sigma_\mathrm{M}^2$) for different values of the sparsity threshold $D$ with $N = 100$, $K = 20$, $\tau_u = K$, $\rho_u = 0$ dB, $T = 250$.}
	\label{fig:MC_power_different_D}
\end{figure}
\begin{figure}[!t]
	\centering
	\subfloat[]{\includegraphics[width=0.48\textwidth]{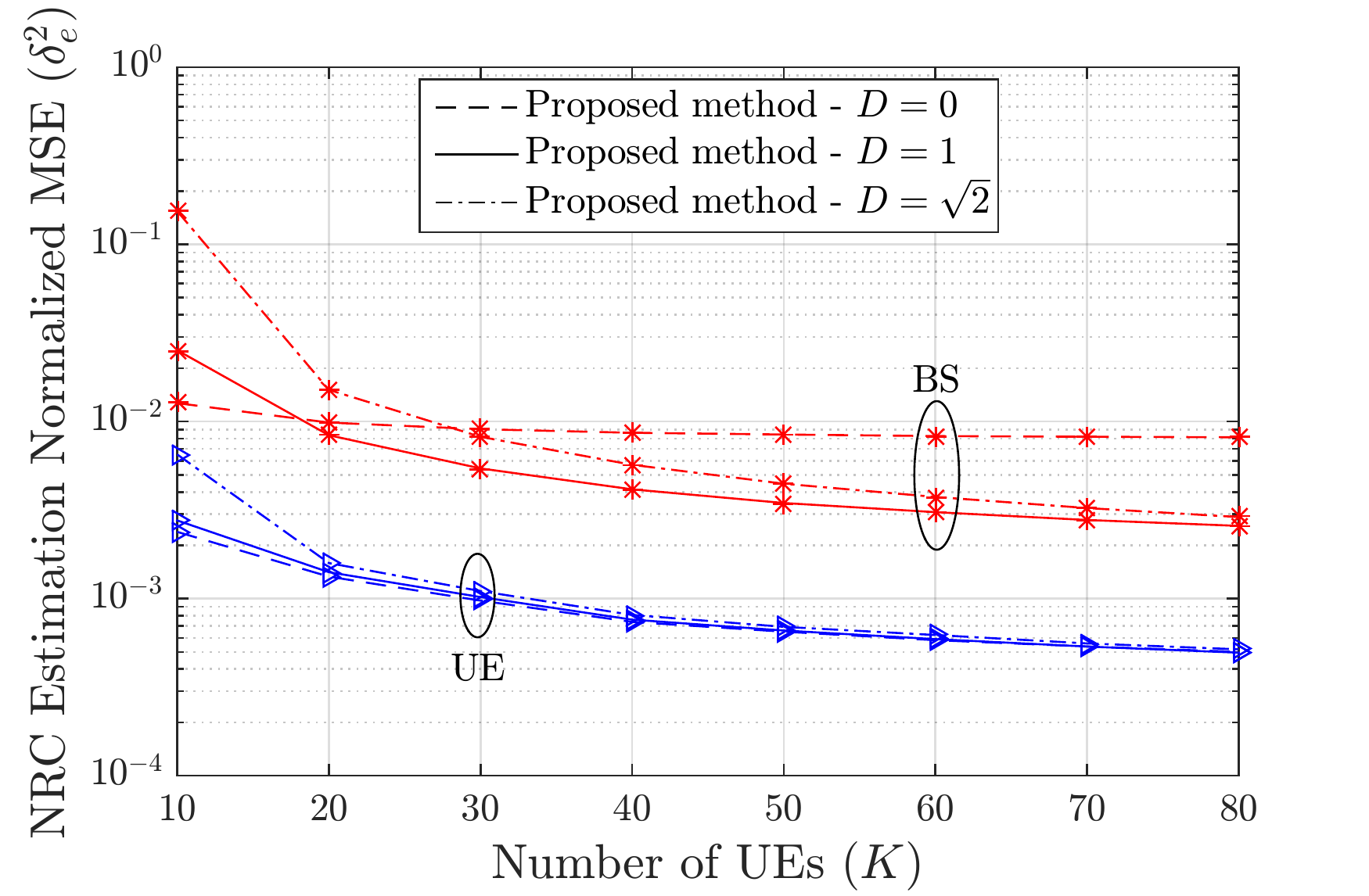}%
		\label{subfig:MSE_K_diffD}}
	\hfil
	\subfloat[]{\includegraphics[width=0.48\textwidth]{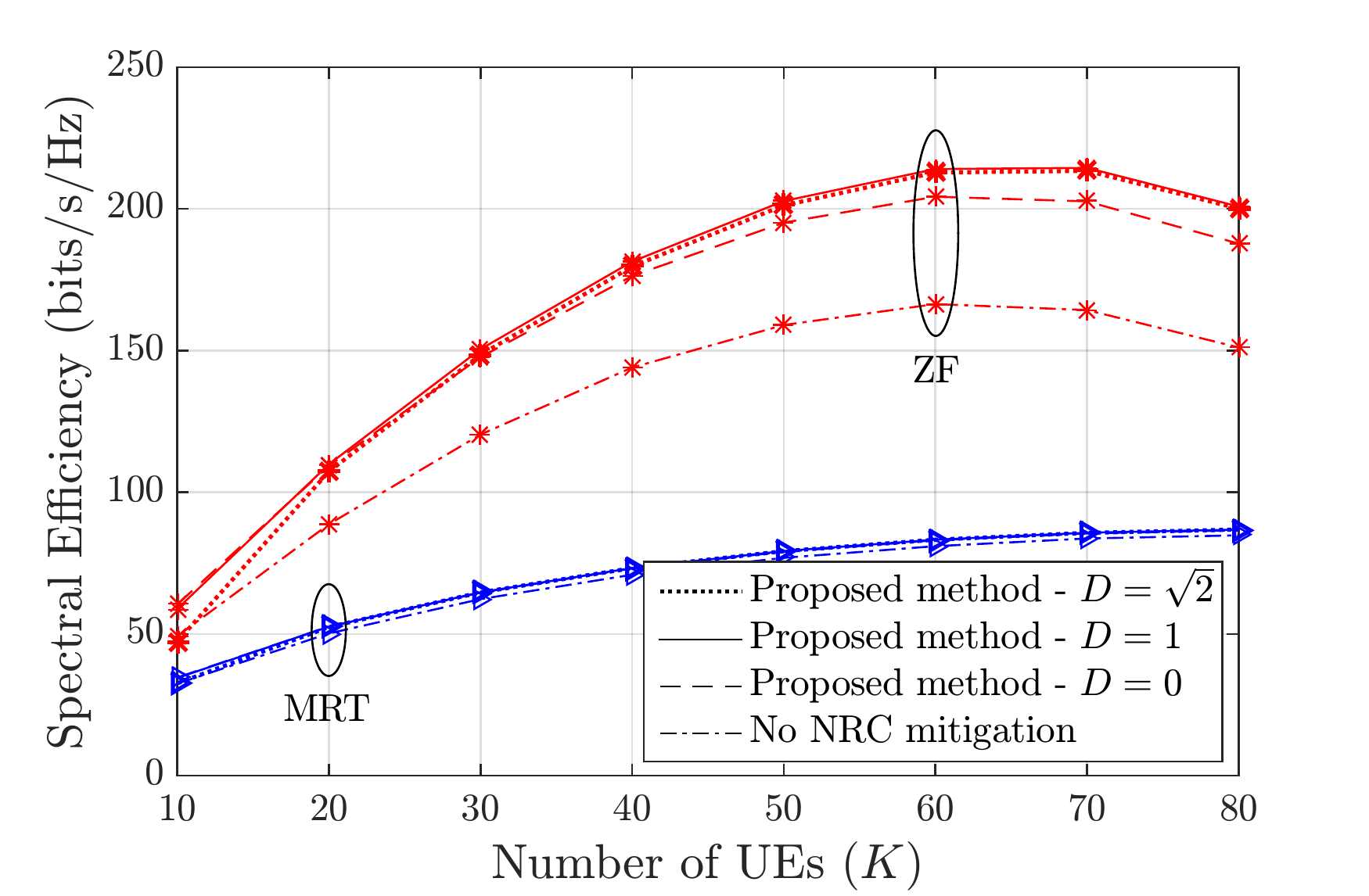}%
		\label{subfig:spec_K_diffD}}
	\caption{\protect\subref{subfig:MSE_K_diffD} NRC estimation normalized MSE and \protect\subref{subfig:spec_K_diffD} system spectral efficiency vs. number of UEs ($K$) for $N = 100$, $\tau_u = K$, $\rho_u = 0$ dB, $T = 250$.}
	\label{fig:K_diffD}
\end{figure}
%As discussed in the previous section, for fixed sparsity threshold $D$, increasing $K$ yields better estimates which in turn yields higher achievable spectral efficiencies. Now, as a complementary study, here we will investigate the spectral efficiency performance with respect to $D$ when $K$ is fixed. In order to find the optimum value of $D$ for different scenarios, in \figurename{~\ref{fig:MC_power_different_D_MSE}} and \figurename{~\ref{fig:MC_power_different_D}}, the effect of $D$ on NRC estimation normalized MSE and spectral efficiency are evaluated against various levels of input reflection coefficients variance, which controls the power of mutual coupling between the BS antennas, for $K = 10$, $20$.
%\figurename{~\ref{fig:MC_power_different_D_MSE}} shows that although having higher mutual coupling power in BS increases the normalized MSE of BS-side NRC estimation, it does not affect the error in estimating UE-side NRC.
%As can be seen in \figurename{~\ref{subfig:MC_K_10}}, in the case where $K = 10$, $D = 0$ which corresponds to the diagonal estimation of BS NRC matrix produces the best result. Whereas, $D = \sqrt{2}$, which correspond to considering the mutual coupling effects of at most $8$ neighboring antennas in our squared antenna layout, is even worse than the case without any NRC mitigation.
Here, we will study the effect of $D$ on the normalized MSE and the system spectral efficiency.
% In this respect, normalized MSE and spectral efficiency are evaluated for different values of $D$ against variance of input reflection coefficient and number of scheduled UEs, in \figurename{~\ref{fig:MC_power_different_D}} and \figurename{~\ref{fig:K_diffD}}, respectively.
In this respect, \figurename{~\ref{subfig:MC_K_20_MSE}} illustrates the normalized MSE of UE and BS NRC estimation under the baseline system settings, with the value of $\sigma_\mathrm{M}^2$ being varied. It can be seen that the choice of $D = 0$, i.e., estimating only the diagonal elements of $\mathbf{B}$, yields the lowest MSE for UE NRC estimation. Note that, in the proposed NRC estimation method, the choice of $D$ influences the UE side estimation as well since $\mathbf{A}$ and $\mathbf{B}$ are estimated iteratively as described in Section \ref{sec:iterative}. On the other hand, the highest BS NRC estimation accuracy is achieved for $D = 0$ only when $\sigma_\mathrm{M}^2 \leq -21$ dB, whereas higher estimation accuracy is obtained for $D = 1$ when $\sigma_\mathrm{M}^2 > -21$ dB. Following that, the spectral efficiencies plotted in \figurename{~\ref{subfig:MC_K_20}} illustrate the combined effect of UE and BS NRC estimation. As can be seen, the highest spectral efficiency is achieved for $D = 0$ when $\sigma_\mathrm{M}^2 \leq -22$ dB and for $D = 1$ when $\sigma_\mathrm{M}^2 > -22$ dB. 
%\figurename{~\ref{subfig:MC_K_20_MSE}} illustrates that for a fixed number of scheduled UEs $K = 20$, the choice of $D = 0$ yields the lowest normalized MSE for UE NRC estimation, while higher choice of $D$ results into higher BS NRC estimation accuracy as the level of NRC grows. The combined effect of BS and UE NRC estimation accuracy can be seen in \figurename{~\ref{subfig:MC_K_20}}, where the spectral efficiency is evaluated for different values of $D$ against the variance of input reflection coefficient for $K = 20$. As can be seen, the diagonal estimation of $\mathbf{B}$, $D = 0$, is the best option only when $\sigma_\mathrm{M}^2 < -22$ dB, and $D = 1$ can be considered as the optimum choice when {$\sigma_\mathrm{M}^2 > -22$ dB}.

For fixed NRC characteristics of $\sigma_\mathrm{L}^2 = \sigma_\mathrm{F}^2 = \sigma_\mathrm{M}^2 = -20$ dB, \figurename{~\ref{fig:K_diffD}} evaluates the normalized estimation MSE and the system spectral efficiency for different values of $D$ and against the number of scheduled UEs $K$. \figurename{~\ref{subfig:MSE_K_diffD}} shows that 
higher UE and BS NRC estimation accuracy is achieved for $D=0$ when $K < 20$, whereas when the number of scheduled users exceeds $K \geq 20$ the choice of $D = 1$ yields the highest BS NRC estimation accuracy. For $K \geq 20$, UE NRC estimation performances are largely similar for all three choices of $D$. Following these, \figurename{~\ref{subfig:spec_K_diffD}} illustrates that from spectral efficiency perspective, the optimum sparsity distance threshold is $D = 0$ for $K < 20$ and $D = 1$ for $K \geq 20$. Thus, in the continuation $D = 0$ and $D = 1$ will be used under the settings of $K < 20$ and $K \geq 20$, respectively.
%while for lower number of UEs $K < 20$, $D = 0$ results into having lower UE and BS NRC estimation MSE, increasing the number of UEs for a fixed NRC level, causes different values of $D$ to have approximately the same UE NRC estimation performance. Thus, for $K \geq 20$, the spectral efficiency is depending on the BS NRC estimation accuracy, where $D = 1$ is shown in \figurename{~\ref{subfig:MSE_K_diffD}} to be the optimum choice. \figurename{~\ref{subfig:spec_K_diffD}} confirms that from spectral efficiency point of view, the best choices of sparsity threshold are $D = 0$ for $K < 20$ and $D = 1$ for $K \geq 20$, and therefore will be used accordingly in the continuation.
As discussed in the previous section, 
when $\mathbf{A} \approx \mathbf{I}_K$, $\mathbf{G}^{*}\mathbf{A}\mathbf{G}^\mathrm{T}$ which is used in the estimation process is of rank $K$. Therefore, having higher number of $K$ increases the accuracy of the BS NRC estimation in the proposed method which facilitates the estimation of more non-diagonal elements in $\mathbf{B}$, i.e., higher values for $D$.

It should be noted that for all the cases in \figurename{~\ref{fig:MC_power_different_D}} and \figurename{~\ref{fig:K_diffD}}, the proposed iterative NRC estimator is executed over a sufficient amount of iterations such that convergence is obtained. This is commonly in the order of $4$ iterations, as illustrated more specifically next.

%\figurename{~\ref{subfig:MSE_K_diffD}} illustrates that, for $K = 10$, the choice of $D = 0$ yields the lowest normalized MSE. Thus, under the given settings, lowest normalized MSE is achieved when the proposed method estimates only the diagonal elements of BS NRC Matrix $\mathbf{B}$.
%\figurename{~\ref{subfig:MC_K_20}} shows that for $K = 20$, the diagonal estimation of $\mathbf{B}$ is the best option only when $\sigma_\mathrm{M}^2 < -22$ dB, while $D = 1$ can be considered as the optimum choice when {$\sigma_\mathrm{M}^2 > -22$ dB}. Based on the results provided in \figurename{~\ref{fig:K_diffD}}, in the continuation, the sparsity threshold is chosen as $D = 0$ for $K < 20$ and $D = 1$ for $K \geq 20$.

\subsubsection{Effect of the Number of Iterations}
%\begin{figure}[!t]
%	\centering
%	\subfloat[]{\includegraphics[width=0.48\textwidth]{iteration_MSE.pdf}%
%		\label{subfig:iter_MSE}}
%	\hfil
%	\subfloat[]{\includegraphics[width=0.48\textwidth]{iteration.pdf}%
%		\label{subfig:iter_spec}}
%	\caption{\protect\subref{subfig:iter_MSE} NRC estimation normalized MSE and \protect\subref{subfig:iter_spec} spectral efficiency vs. NRC estimation iteration number for $N = 100$, $K = 20$, $\tau_u = K$, $\rho_u = 0$ dB, $T = 250$.}
%	\label{fig:iteration}
%\end{figure}
\begin{figure}[!t]
	\centering
	\includegraphics[width=0.48\textwidth]{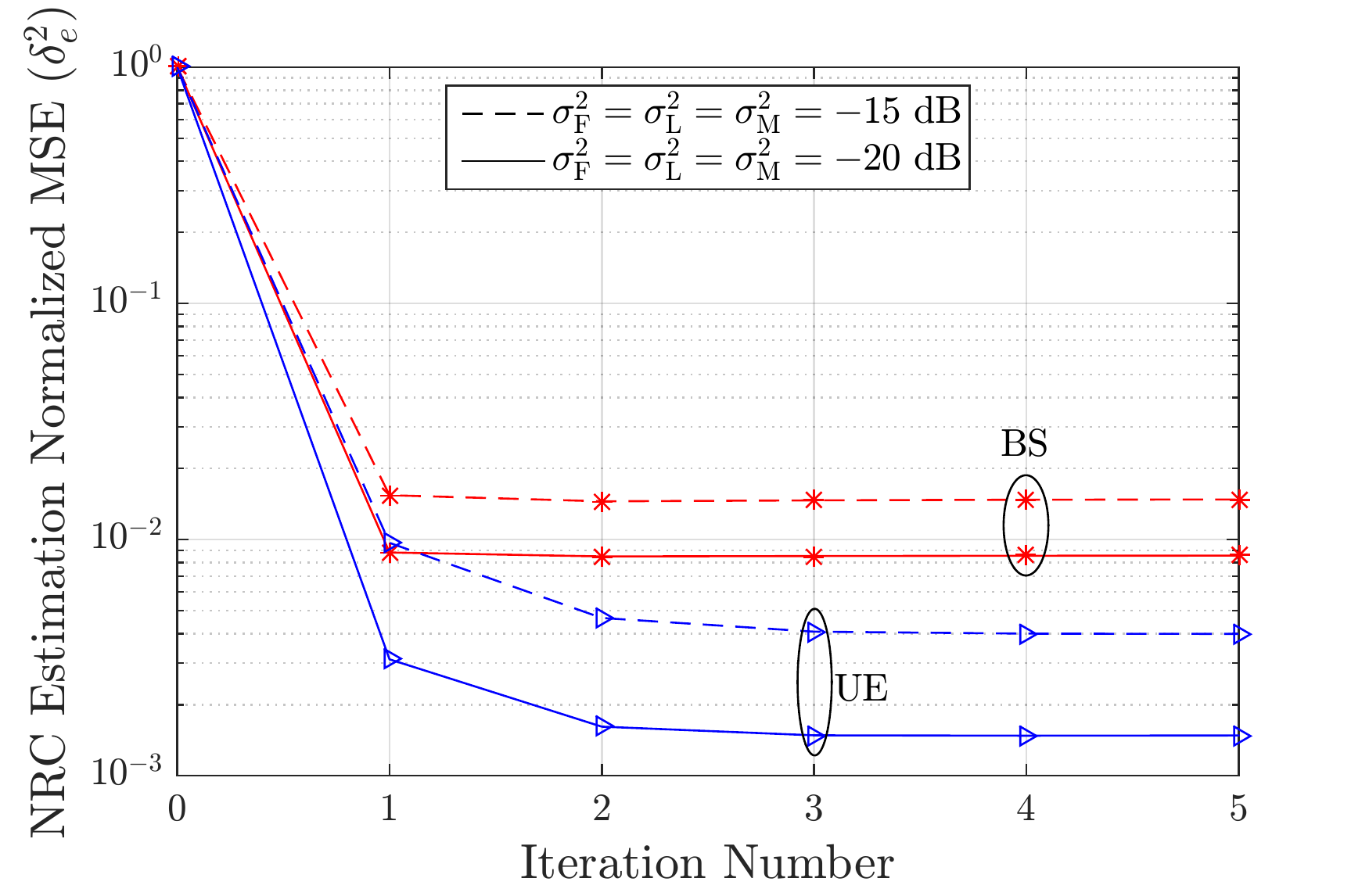}
	\caption{NRC estimation normalized MSE vs. NRC estimation iteration number for $N = 100$, $K = 20$, $\tau_u = K$, $\rho_u = 0$ dB, $T = 250$.}
	\label{fig:iteration}
\end{figure}
%\begin{figure}[!t]
%	\centering
%	\includegraphics[width=0.48\textwidth]{iteration.pdf}
%	\caption{Spectral efficiency vs. NRC estimation iteration number for $N = 100$, $K = 20$, $\tau_u = K$, $\rho_u = 0$ dB, $T = 250$.}
%	\label{fig:iteration}
%\end{figure}
\figurename{~\ref{fig:iteration}} illustrates the reduction in NRC estimation normalized MSE over NRC estimation iteration steps. %As can be seen, 
%the proposed NRC estimator converges to its optimum performance only after $2$ rounds of iterations when MRT precoding is employed. The convergence requires more iteration rounds for higher NRC levels and for ZF precoding. However, 
%the BS-side NRC estimator converges faster than the UE-side NRC estimator.
It is shown in \figurename{~\ref{fig:iteration}} that, even with high NRC levels of $\sigma_\mathrm{L}^2 = \sigma_\mathrm{F}^2 = \sigma_\mathrm{M}^2 = -15$ dB, having $4$ iteration rounds is sufficiently good for the proposed NRC estimator to converge. Therefore, in the continuation, we set the number of iteration rounds to $4$.

\subsubsection{Effect of Number of Scheduled Users $K$}
\begin{figure}[!t]
	\centering
	\subfloat[]{\includegraphics[width=0.48\textwidth]{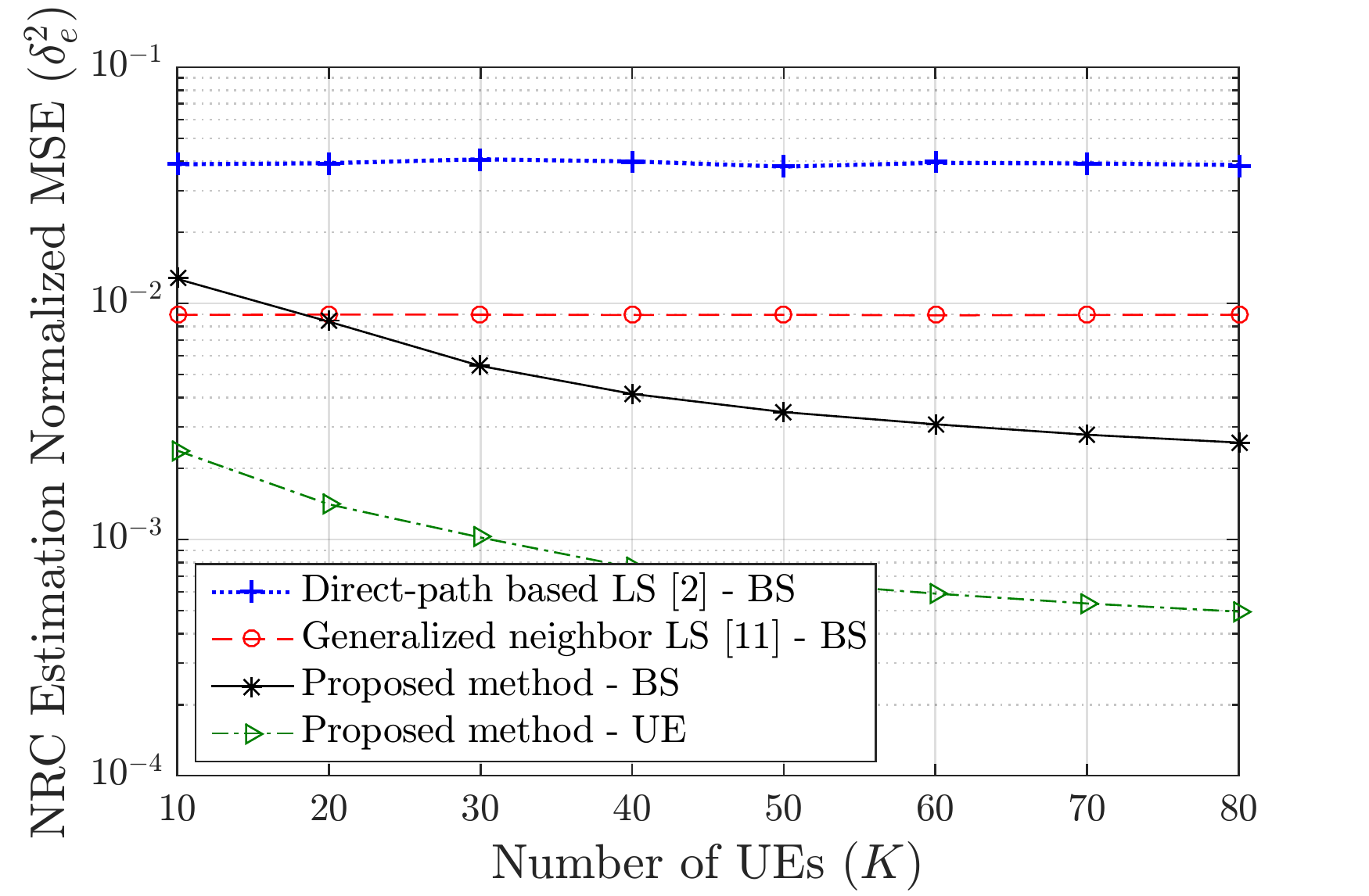}%
		\label{subfig:K_MSE}}
	\hfil
	\subfloat[]{\includegraphics[width=0.48\textwidth]{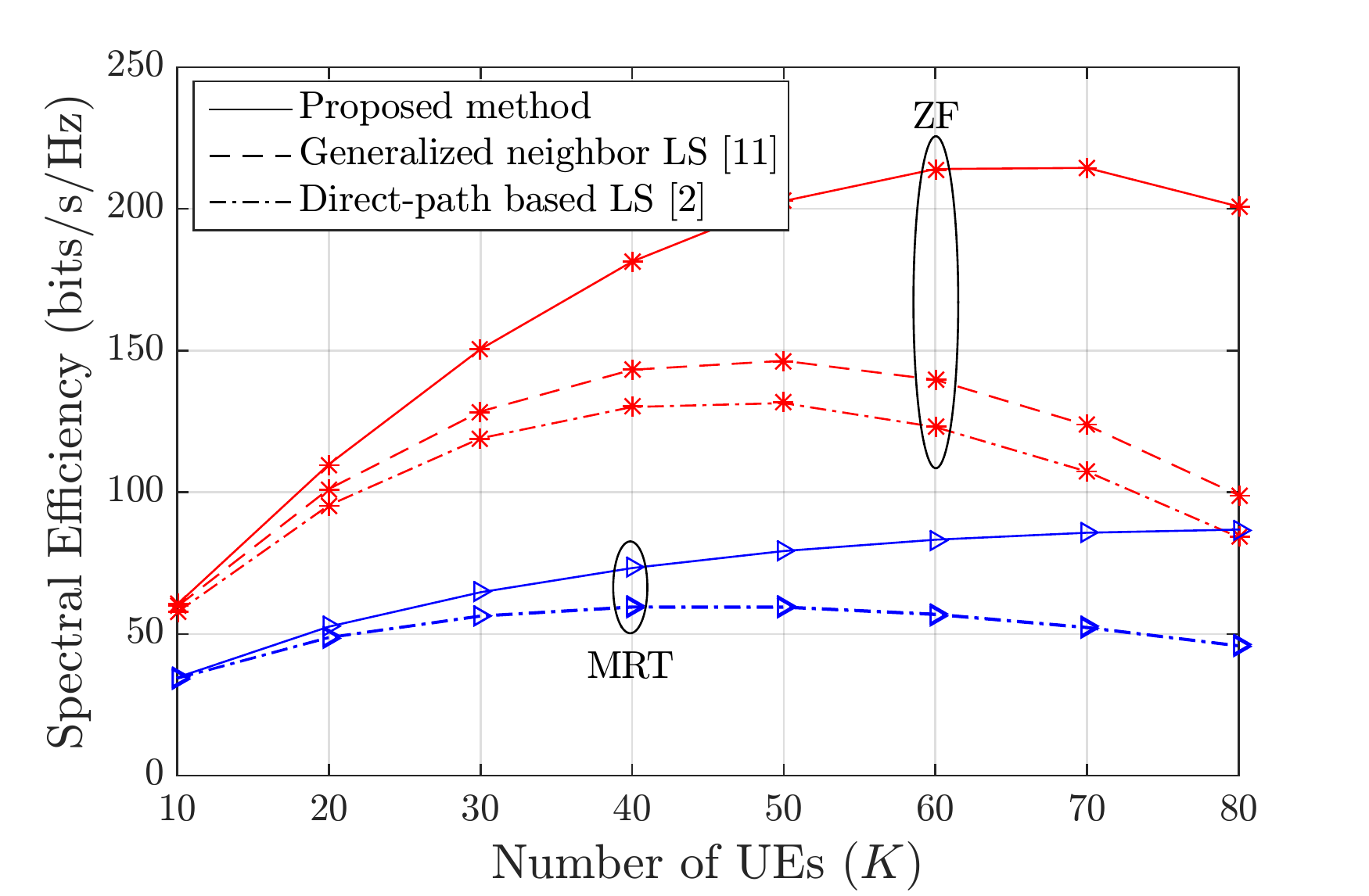}%
		\label{subfig:K}}
	\caption{\protect\subref{subfig:K_MSE} NRC estimation normalized MSE and \protect\subref{subfig:K} system spectral efficiency vs. number of UEs ($K$) for $N = 100$, $\tau_u = K$, $\rho_u = 0$ dB, $T = 250$.}
	\label{fig:K}
\end{figure}
%\begin{figure}[!t]
%	\centering
%	\includegraphics[width=0.48\textwidth]{K_MSE.pdf}
%	\caption{NRC estimation normalized MSE vs. number of UEs ($K$) for $N = 100$, $\tau_u = K$, $\rho_u = 0$ dB, $T = 250$.}
%	\label{fig:K_MSE}
%\end{figure}
%\begin{figure}[!t]
%	\centering
%	\includegraphics[width=0.48\textwidth]{K.pdf}
%	\caption{Spectral efficiency vs. number of UEs ($K$) for $N = 100$, $\tau_u = K$, $\rho_u = 0$ dB, $T = 250$.}
%	\label{fig:K}
%\end{figure}
In \figurename{~\ref{fig:K}}, the NRC estimation normalized MSE and the system spectral efficiency are plotted against the number of scheduled UEs $K$ for $\sigma_\mathrm{M}^2 = -20$ dB.
\figurename{~\ref{subfig:K_MSE}} shows that while direct-path based LS has the worst performance, the proposed method is the best option for estimating BS NRC for $K \geq 20$ with a high accuracy where MSE is in the order of $10^{-3}$. For direct-path based LS \cite{shepard_argos:_2012} and generalized neighbor LS \cite{vieira_reciprocity_2014}, the normalized MSE for UE side NRC is not shown. It is mentioned in \cite{shepard_argos:_2012} and \cite{vieira_reciprocity_2014} that additional downlink pilot signaling per coherence interval can be used together with UE side estimation for UE side NRC acquisition. However, no detailed description is provided on the actual pilot signal structure or the actual estimation method. %One significant drawback of such downlink pilots based method is that part of the available resources are wasted whereas no additional pilot signaling is required for the proposed estimation scheme per coherence interval. 

%As can be seen in \figurename{~\ref{fig:K}}
%in the zoomed inner plots, even for the lowest number of scheduled UEs $K = 10$, when the proposed NRC estimator assumes that BS NRC matrix is diagonal $D = 0$, 
%the proposed NRC estimation and mitigation scheme outperforms direct-path based LS and generalized neighbor LS methods.
The corresponding system spectral efficiency performances are evaluated and shown in \figurename{~\ref{subfig:K}}. The proposed NRC estimation and mitigation scheme clearly outperforms the direct-path based LS and generalized neighbor LS methods.
The difference between the performance of the proposed method and the other two methods increases as $K$ grows. Remarkably, for $K = 70$, the difference between the proposed method and the other two methods is already in the order of $100$ bits/s/Hz. %The reason is that while the two LS based methods require to spend larger portion of each coherence interval for downlink pilot transmission for higher number of scheduled UEs, the proposed method gets more accurate as the ratio $K/R_j$ increases. It should be noted that for $K > 20$ we could have used $D > 1$ which would have further improved the performance of the proposed method.
%Note that the proposed method could have been further optimized by setting  $D > 1$ for $K > 20$, thus the difference would be even higher.
Another advantage in utilizing the proposed NRC estimation scheme is that the optimum number of UEs $K_{opt}$, which is defined as the number of scheduled UEs which maximizes the spectral efficiency, is higher compared to the other two NRC estimation methods. For instance under ZF precoding $K_{opt}$ is between $60$ and $70$ for the proposed method whereas for LS based methods $K_{opt}$ is only around $50$.

\subsubsection{Effect of Downlink SNR}
\begin{figure}[!t]
	\centering
	\includegraphics[width=0.48\textwidth]{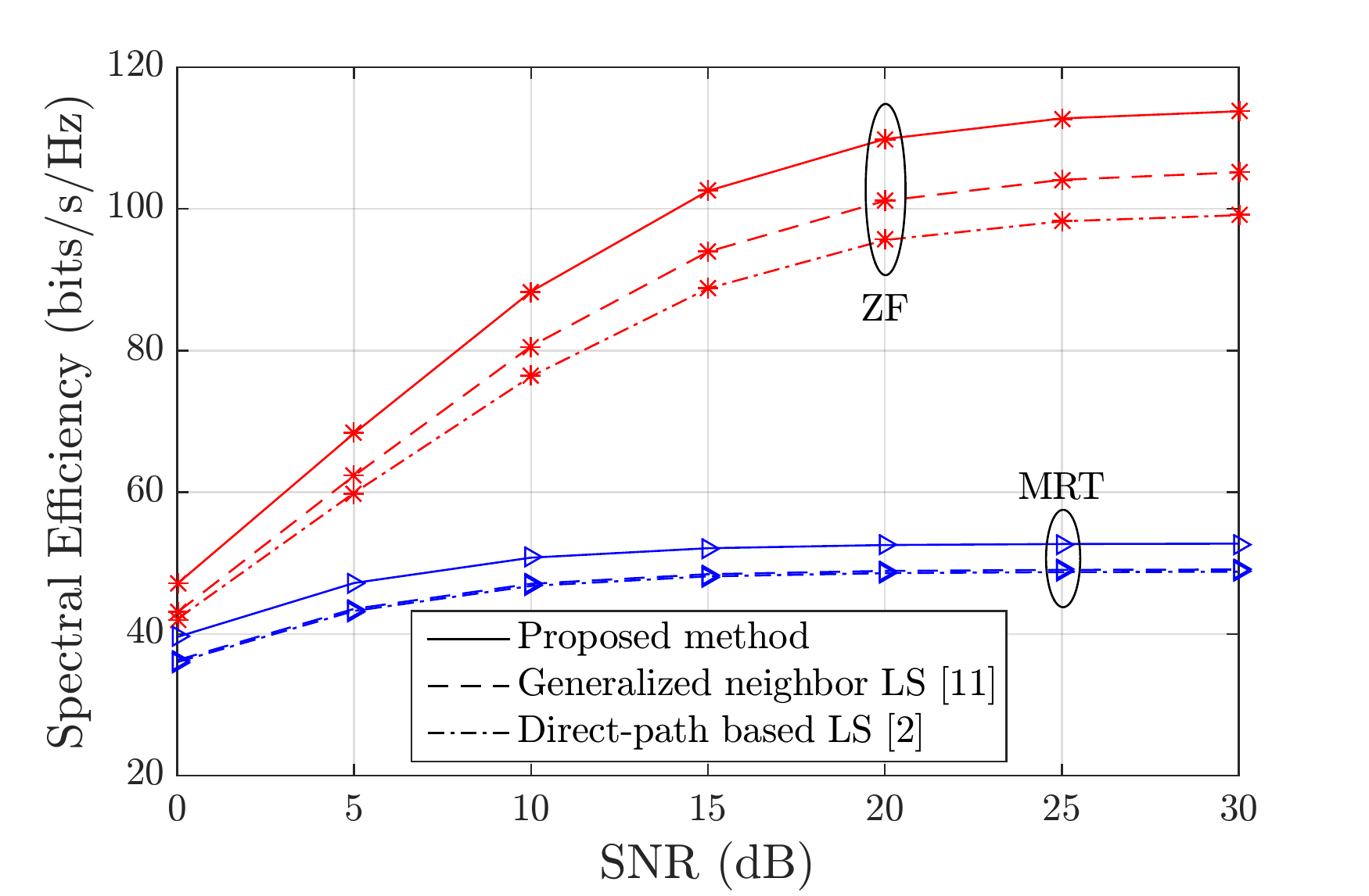}
	\caption{Spectral efficiency vs. downlink SNR ($\rho_d$) for $N = 100$, $K = 20$, $\tau_u = K$, $\rho_u = 0$ dB, $T = 250$.}
	\label{fig:SNR}
\end{figure}
In \figurename{~\ref{fig:SNR}}, the system spectral efficiency is plotted against the downlink SNR when $\sigma_\mathrm{M}^2 = -20$ dB. The results show clear advantage in employing the proposed method in estimating NRC for all SNR values. The proposed estimation and mitigation method outperforms the LS based methods for both low and high SNR regions. Especially, the performance difference is most visible for high SNR region under ZF precoding.
%For low SNR region, the performance of all the methods are mainly limited by noise and the difference between the proposed method and the other two is mostly due to not spending time on sending downlink pilots. On the other hand, as the SNR increases, the difference between the performance of the proposed method and that of other methods gets higher, mainly for ZF precoding which is affected more by NRC, which is due to its higher accuracy in NRC estimation.

\subsubsection{Effect of Input Reflection Coefficient}
\begin{figure}[!t]
	\centering
	\includegraphics[width=0.48\textwidth]{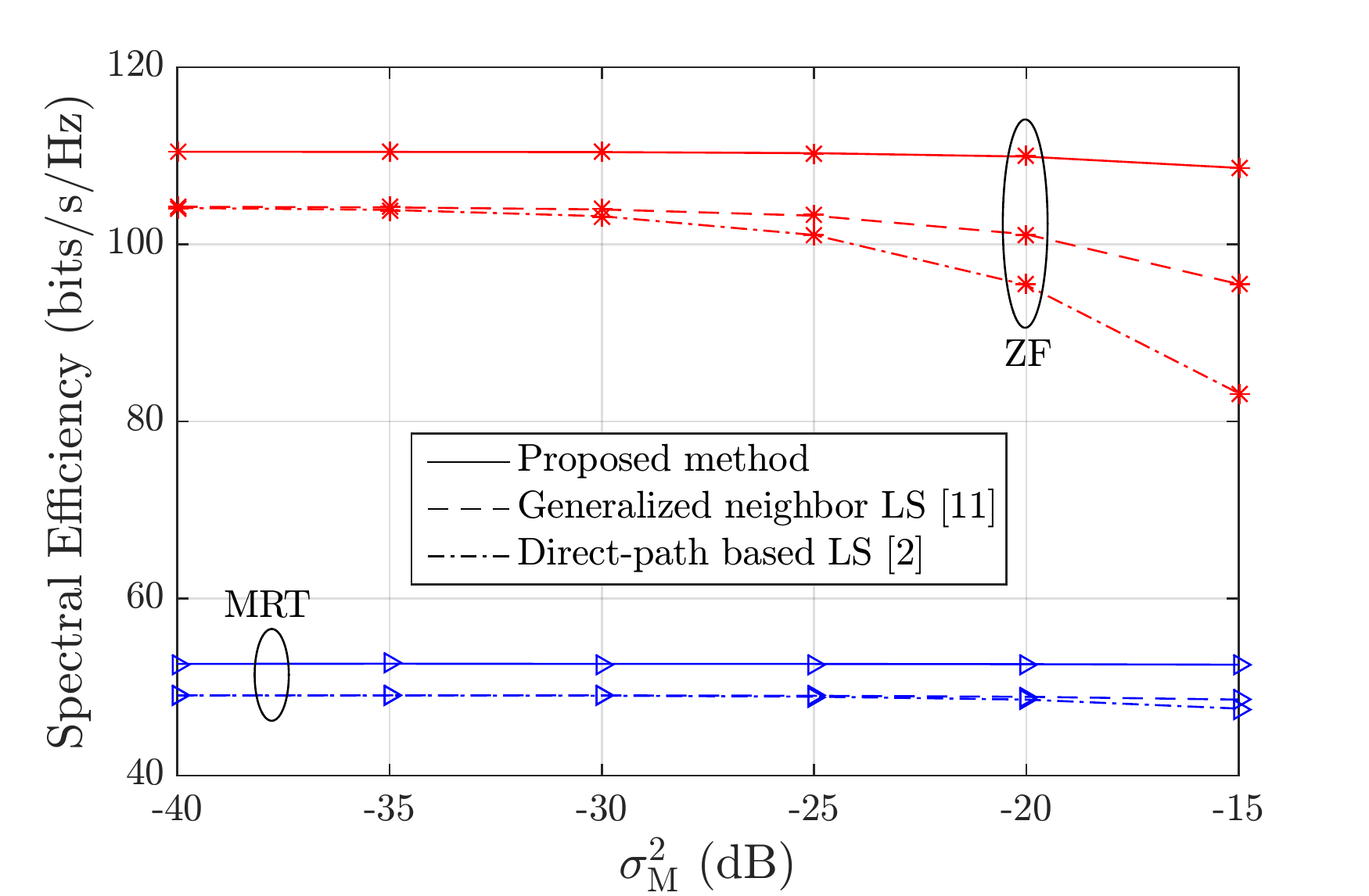}
	\caption{Spectral efficiency vs. input reflection coefficients variance ($\sigma_\mathrm{M}^2$) for $N = 100$, $K = 20$, $\tau_u = K$, $\rho_u = 0$ dB, $T = 250$.}
	\label{fig:MC_power}
\end{figure}
\figurename{~\ref{fig:MC_power}} shows the impact of the variance of the input reflection coefficients on the achievable spectral efficiency. The proposed estimation and mitigation method again outperforms the other two LS based methods. %As can be seen, regardless of the power of mutual coupling between the BS antennas, the proposed method outperforms others.
The difference between the proposed method and the other two methods increases as $\sigma_\mathrm{M}^2$ grows, which is due to the ability of the proposed method to estimate the non-diagonal elements in BS NRC matrix. It should be noted that $D = 1$ is used for obtaining the results in \figurename{~\ref{fig:MC_power}}, and there is still room for improving the performance of the proposed method by adaptively selecting the optimum $D$ according to the level of $\sigma_\mathrm{M}^2$ as shown already in \figurename{~\ref{subfig:MC_K_20}}.

\subsubsection{Summary of the Obtained Results}
Overall, as observed through extensive numerical evaluations in various scenarios, the proposed NRC estimation method outperforms the other two state-of-the-art methods. Selected technical aspects can be summarized as follows:
\begin{itemize}
	\item Employing the proposed NRC estimation method eliminates the need to send downlink demodulation pilots since the proposed OTA framework facilitates estimating both the BS side and UE side NRC characteristics in the base station. Therefore, more time-frequency resources can be allocated in each coherence interval for actual downlink data transmission purposes which improves the spectral efficiency.
	%\item Number of iteration rounds for estimated NRC convergence is low, at most $4$, which guaranties fast and computationally feasible convergence.
	\item The proposed NRC estimation method is more and more superior over the two reference methods when the number of scheduled UEs $K$ grows. The reason is that increasing $K$ is forcing the other two NRC estimation methods to spend more time for downlink pilot transmission in each coherence interval, while a larger number of scheduled users improves the accuracy of the proposed NRC estimation method.
	\item Due to the ability to estimate also non-diagonal elements of the BS NRC matrix, the difference between the performance of the proposed NRC estimation method and the other two methods increases as the power of BS antenna mutual coupling mismatch grows.
\end{itemize}

\section{Conclusion} \label{sec:conclusions}
In this work, we proposed an efficient NRC estimation and mitigation framework for multi-user massive MIMO TDD networks to compensate the jointly coexisting BS and UE side NRC. In general, even relatively modest NRC levels can cause significant performance loss in the achievable spectral efficiency when only standard NRC-blind MRT or ZF downlink precoding is employed. A novel OTA-based approach incorporating a dedicated round-trip pilot signaling with small pilot overhead together with sparsity-aided efficient iterative estimation techniques were proposed for the acquisition of NRC matrices at BS. Unlike the existing state-of-the-art methods, the proposed NRC estimation method acquires both the UE transceiver NRC as well as the BS transceiver NRC, and does not rely on downlink pilot transmission during the actual data transmission phase to compensate the NRC in the UE side. Therefore, it can be efficiently employed in massive MIMO systems that rely only on the statistical knowledge of the beamformed downlink channels at terminals for data decoding with very low system pilot overhead. The extensive computer simulations showed that for practical values of the NRC levels, SNRs and the number of spatially multiplexed users, the proposed estimation and mitigation method clearly outperforms the existing state-of-the-art methods in terms of the system spectral efficiency.

\appendix[Proof for estimation of $\mathbf{A}$] \label{App:UE}
Let
\begin{equation}
\mathcal{L} \overset{\Delta}{=} \left|\left| \mathbf{Q}-\sqrt{\tilde{\rho}_{u}}\sqrt{\tilde{\rho}_{d}}\hat{\mathbf{G}}^*{\mathbf{A}} \hat{\mathbf{G}}^\mathrm{T} \hat{\mathbf{B}}(m)\right|\right|^2_F.
\end{equation}
Then,
\begin{equation}\label{eq:app1}
\begin{aligned}
\mathcal{L} &= \sum_{j=1}^{N} \left|\left| \mathbf{q}_j-\sqrt{\tilde{\rho}_{u}}\sqrt{\tilde{\rho}_{d}}\hat{\mathbf{G}}^*{\mathbf{A}} \left[\hat{\mathbf{G}}^\mathrm{T}\hat{\mathbf{B}}(m)\right]_j\right|\right|^2 \\
&= \sum_{j=1}^{N} \left|\left| \mathbf{q}_j-\sqrt{\tilde{\rho}_{u}}\sqrt{\tilde{\rho}_{d}}\hat{\mathbf{G}}^* \mathrm{diag}\left(\left[\hat{\mathbf{G}}^\mathrm{T}\hat{\mathbf{B}}(m)\right]_j\right)\boldsymbol{\xi}\right|\right|^2,
\end{aligned}
\end{equation}
where $\boldsymbol{\xi} \overset{\Delta}{=} \left[a_1, a_2, \cdots, a_K\right]^\mathrm{T}$.

By using $\mathbf{W}_j=\sqrt{\tilde{\rho}_{u}}\sqrt{\tilde{\rho}_{d}}\hat{\mathbf{G}}^*\mathrm{diag}\left(\left[\hat{\mathbf{G}}^\mathrm{T}\hat{\mathbf{B}}(m)\right]_j\right)$, \eqref{eq:app1} can be re-written as
\begin{equation}
\begin{aligned}
\mathcal{L} &= \sum_{j=1}^{N} \left|\left| \mathbf{q}_j-\mathbf{W}_j\boldsymbol{\xi}\right|\right|^2 \\
&= \sum_{j=1}^{N} \left|\left| \bar{\mathbf{q}}_j-\bar{\mathbf{W}}_j\boldsymbol{\psi}\right|\right|^2 \overset{\Delta}{=} \bar{\mathcal{L}}\left(\boldsymbol{\psi}\right),
\end{aligned}
\end{equation}
where $\boldsymbol{\psi} \overset{\Delta}{=} \left[\Re\{\boldsymbol{\xi}^\mathrm{T}\},\Im\{\boldsymbol{\xi}^\mathrm{T}\}\right]^\mathrm{T}$. Therefore, the solution $\hat{\mathbf{A}}(m)$ can be obtained by
\begin{equation}
\hat{\mathbf{A}}(m)= \mathrm{diag}\left(\left[\mathbf{I}_K, i\mathbf{I}_K\right]\hat{\boldsymbol{\psi}}\right),
\end{equation} 
where 
\begin{equation} \label{eq:convex}
\hat{\boldsymbol{\psi}} = \underset{\boldsymbol{\psi} \in \mathbb{R}^K}{\text{argmin}} \ \bar{\mathcal{L}}\left(\boldsymbol{\psi}\right).
\end{equation} 
Since $\bar{\mathcal{L}}\left(\boldsymbol{\psi}\right)$ is convex, \eqref{eq:convex} can be solved from the partial derivative equation $\frac{\partial \bar{\mathcal{L}}\left(\boldsymbol{\psi}\right)}{\partial \boldsymbol{\psi}} = 0$, which finally yields the solution given by \eqref{eq:UE NRC solution}. %\footnote{The estimation of $\mathbf{A}$ is based on a work presented in \cite{ngo_evd-based_2012}. The proof was omitted in \cite{ngo_evd-based_2012} due to space constraints and is provided based on a discussion with the authors.}.
%\vspace{5mm}
%
%\footnotesize
%$^*$The estimation of $\mathbf{A}$ is based on a work presented in \cite{ngo_evd-based_2012}. The proof was omitted in \cite{ngo_evd-based_2012} due to space constraints and is provided based on a discussion with the authors.
%
%\normalsize

\section*{Acknowledgment}
The estimation of $\mathbf{A}$ is building on the work presented in \cite{ngo_evd-based_2012}. The proof in Appendix A is a result of discussion with the authors of \cite{ngo_evd-based_2012}, H. Q. Ngo and E. G. Larsson, whose technical guidance is greatly acknowledged.

% or


%\appendices
%% you can choose not to have a title for an appendix
%% if you want by leaving the argument blank
%\section{}

%\section*{Acknowledgment}

\ifCLASSOPTIONcaptionsoff
  \newpage
\fi



\bibliographystyle{IEEEtran}
\bibliography{IEEEabrv,NRC_MAMI_ref}
\end{document}